\documentclass{aa}
\usepackage{amsmath} 
\usepackage{amssymb} 
\usepackage{xcolor}
\usepackage[]{geometry}
\usepackage{txfonts}
\usepackage{verbatim}
\usepackage{enumitem}
\usepackage{epsfig}
\usepackage{multirow}
\usepackage{tablefootnote}
\usepackage{stfloats}
\usepackage{subcaption}
\usepackage{booktabs}
\usepackage{adjustbox}

\def\erosita{{\small eROSITA}}
\def\rosat{{\small ROSAT}}
\def\xmm{{XMM-Newton}}

\newcommand{\degree}{\ensuremath{^\circ}}

\def\be{\begin{equation}}
\def\ee{\end{equation}}

\usepackage{listings}

\begin{document}
\titlerunning{eROSITA broadband maps and comparison with ROSAT}
\authorrunning{Zheng et al.}

\title{Broadband maps of eROSITA and their comparison with the ROSAT survey}
\subtitle{}
\author{
Xueying Zheng \inst{1},
Gabriele Ponti \inst{2,1},
Michael Freyberg \inst{1},
Jeremy Sanders \inst{1},
Nicola Locatelli \inst{2,1} \thanks{nicola.locatelli@inaf.it},
Andrea Merloni \inst{1},
Andy Strong \inst{1},
Manami Sasaki \inst{3},
Johan Comparat \inst{1},
Werner Becker \inst{1,4},
Juergen Kerp\inst{5},
Chandreyee Maitra \inst{1},
Teng Liu \inst{1},
Peter Predehl \inst{1},
Konstantina Anastasopoulou \inst{2},
Georg Lamer \inst{6}
}

\institute{
     Max-Planck-Institut f{\"u}r extraterrestrische Physik, Gießenbachstraße 1, D-85748 Garching bei M\"unchen , Germany
\and INAF-Osservatorio Astronomico di Brera, Via E. Bianchi 46, I-23807 Merate (LC), Italy
\and Dr. Karl Remeis Observatory, Erlangen Centre for Astroparticle Physics, Friedrich-Alexander-Universit\"at Erlangen-N\"urnberg Sternwartstraße 7, 96049 Bamberg, Germany
\and Max-Planck-Institut f{\"u}r Radioastronomie, Auf dem H\"ugel 69, 53121 Bonn, Germany 
\and Argelander-Institut f{\"u}r Astronomie (AIfA), University of Bonn, Auf dem H\"ugel 71, 53121 Bonn, Germany
\and Leibniz-Institut f{\"u}r Astrophysik, An der Sternwarte 16, 14482 Potsdam, Germany}

    \date{}
    \abstract{
By June of 2020, the extended ROentgen Survey with an Imaging Telescope Array (\erosita{}) on board the Spectrum Roentgen Gamma observatory had completed its first of the planned eight X-ray all-sky survey (eRASS1). The large effective area of the X-ray telescope makes it ideal for a survey of the faint X-ray diffuse emission over half of the sky with an unprecedented energy resolution and position accuracy.
In this work, we produce the X-ray diffuse emission maps of the eRASS1 data with a current calibration, covering the energy range from 0.2 to 8.0 keV. We validated these maps by comparison with X-ray background maps derived from the \rosat{} All Sky Survey (RASS).
We generated X-ray images with a pixel area of 9 arcmin$^2$ using the observations available to the German \erosita{} consortium. The contribution of the particle background to the photons was subtracted from the final maps. 
We also subtracted all the point sources above a flux threshold dependent on the goal of the subtraction, exploiting the eRASS1 catalog that will soon be available.
The accuracy of the eRASS1 maps is shown by a flux match to the RASS X-ray maps, obtained by converting the \erosita{} rates into equivalent \rosat{} count rates in the standard \rosat{} energy bands R4, R5, R6, and R7, within 1.25$\sigma$. We find small residual deviations in the R4, R5, and R6 bands, where \erosita{} tends to observe lower flux than \rosat{} ($\sim$ 11\%), while a better agreement is achieved in the R7 band ($\sim$ 1\%). The eRASS maps exhibit lower noise levels than RASS maps at the same resolution above 0.3 keV.
We report the average surface brightness and total flux of different large sky regions as a reference.
The detection of faint emission from diffuse hot gas in the Milky Way is corroborated by the consistency of the eRASS1 and RASS maps shown in this paper and by their comparable flux dynamic range.}

\maketitle

\section{Introduction}\label{sec:intro}

The soft X-ray images provided by the \rosat{} All Sky Survey (RASS) \citep{Snowden1997ApJ} are one of the most important legacies of the \rosat\ mission. 
They have revealed the morphology of the X-ray diffuse emission in the 0.1--2.1 keV with high angular resolution and low instrumental background \citep{Snowden1997ApJ}.
The diffuse X-ray emission detected in RASS is primarily attributed to Galactic hot rarefied plasma ($T> 10^6$ K) and its components: a close and unabsorbed thermal plasma filling the Local Bubble \citep{Lallement2018AA, Lallement2015MmSAI} (called the Local Hot Bubble, LHB); an absorbed thermal plasma from the Milky Way circum-Galactic medium \citep[CGM,][]{Tumlinson2017ARAA,Kerp1999AA, Pietz1998AA};  very close emission from charge-exchange processes triggered by the solar wind (i.e., the solar wind charge exchange; SWCX). In addition to these the contributions of localized supernova remnants and clusters \citep{Snowden1993ApJ, McCammon2002ApJ} and the large number of unresolved extragalactic sources (mostly active galactic nuclei; AGN) forming the cosmic X-ray background (CXB) complete the census and explain the total amount of the diffuse emission observed in RASS.

\begin{figure}[b]
    \centering
    \includegraphics[width=0.45\textwidth, trim=5 25 20 30,clip]{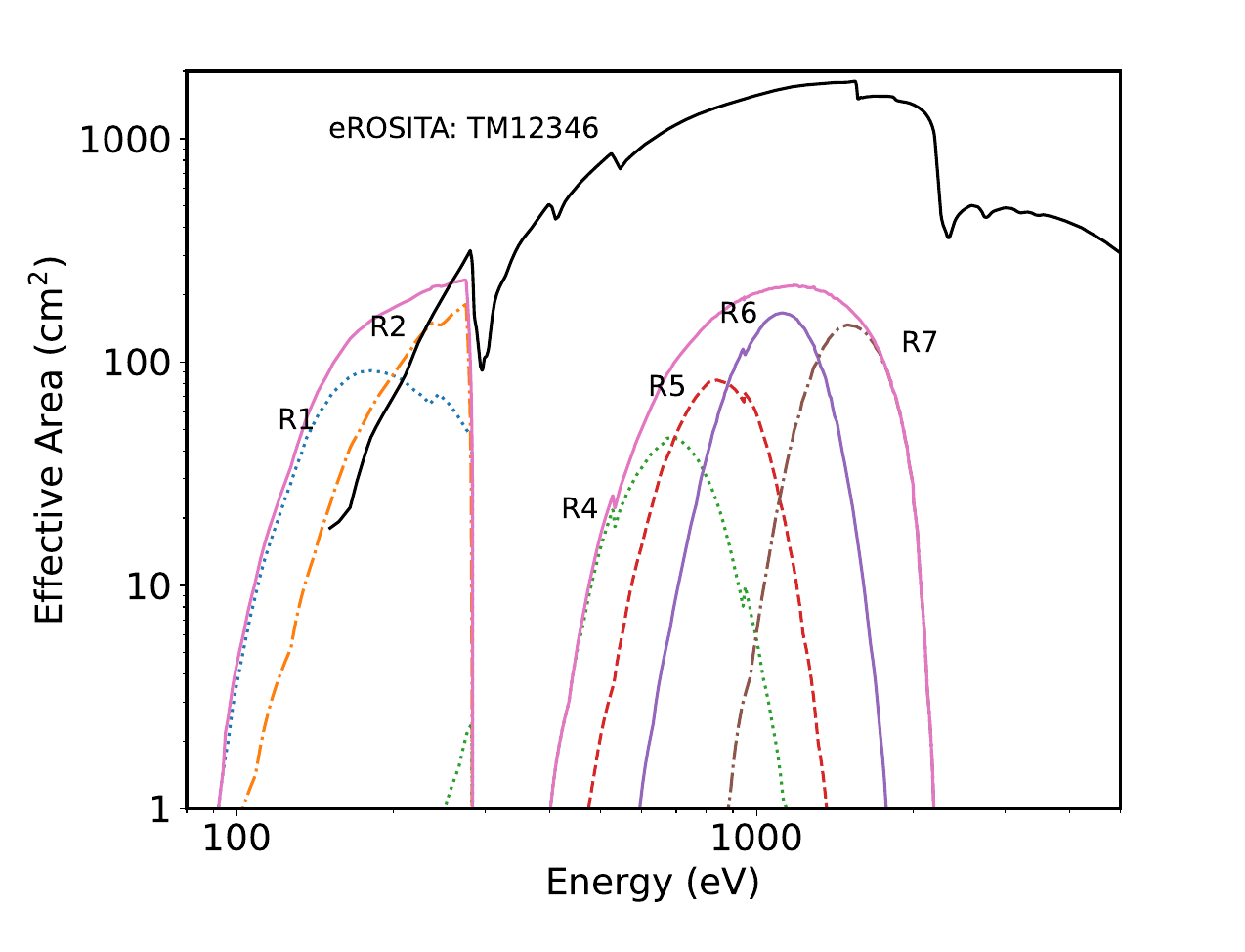}
    \caption{Effective area curves of \rosat{} and \erosita{}. The global \rosat{} effective area (solid pink line) is composed of six standard bands (labeled). The \erosita{} broadband effective area curve (solid black line) combines cameras TM 1, 2, 3, 4, and 6.} 
    \label{fig:effective_area}
    \vspace{-1pt}
\end{figure}
The technological improvements associated with the recently launched extended ROentgen Survey with an Imaging Telescope Array \citep[\erosita{} hereafter, ][]{Predehl2021AA,erositabook} have recently provided the opportunity to explore the faint diffuse emission over the whole sky in greater detail. 
\erosita{} aims to complete eight All-Sky Surveys (called eRASS), in order to reach a sensitivity that is about 20 times greater than that of RASS based on the \erosita{} large effective area (it is about one order of magnitude larger than \rosat{} above 0.3 keV; see Fig.\ref{fig:effective_area}) combined with its $\rm \sim deg^2$ field of view. 
The sensitive energy band of \erosita{} ranges from 0.2 to 8.0 keV.

The \erosita{} maps are not expected to be identical to the RASS maps mostly because of the different instrumental background, space environment, optics, and detector design that characterize the two instruments.
The \erosita{} telescope incorporates seven PN-CCD cameras (an enhanced type of the \xmm\ EPIC-pn focal plane detector), while \rosat{} is equipped with proportional counters (PSPC).
An incomplete but instructive list of examples includes that the \rosat{} PSPC is blind against optical light by construction, whereas the detectors of \erosita{} are sensitive to it; and
\erosita\ orbits the L2 point, while \rosat\ was located in a low Earth orbit \citep{Sunyaev2021AA, 2020Predehl_instr,Truemper1982AdSpR}, leading to a different injection rate of solar induced X-ray and micro-meteorite impact. 
By comparing the observations, we aim to assess potential differences in the scientific output of the two surveys. In addition, the consistency between the sky fluxes will instead validate the calibrations of both instruments.

\begin{table}
    \centering
\begin{tabular}{ccc}
\hline Band (Combined) & Band & Range $\mathrm{keV}$\\
\hline \multirow{2}{*}{R12 ($\frac{1}{4}~\mathrm{keV}$)} & R1  & $0.11-0.284$\\
\multirow{2}{*}{} & R2  & $0.14-0.284$\\
\hline \multirow{2}{*}{R45 ($\frac{3}{4}~\mathrm{keV}$)} & R4  & $0.44-1.01$ \\
\multirow{2}{*}{} & R5 & $0.56-1.21$ \\
\hline \multirow{2}{*}{R67 ($1.5~\mathrm{keV}$)} & R6  & $0.73-1.56$ \\
\multirow{2}{*}{} & R7  & $1.05-2.04$\\     
\hline
\end{tabular}
    \caption{The ROSAT standard band energy ranges. We redefined the band boundaries at 10\% of the peak response in the band.}
    \label{tab:ro_bands}
\end{table}
In addition, as a very important output, we present the soft broadband maps based on the data from the \erosita{} first all-sky survey  (eRASS1, conducted from 11 December 2019 to 11 June 2020). We validate the \erosita\ maps by performing a quantitative comparison of the emission observed in RASS in the \rosat\ standard bands R4, R5, R6, and R7 (see Tab. \ref{tab:ro_bands} for a description of the bands).

The paper is structured in the following way: Section \ref{sec:process} outlines the data reduction process we used to produce the eRASS1 X-ray diffuse maps, including the corrections for exposure, vignette, point sources, and instrumental background. Section \ref{sec:broadmap} presents the eRASS1 broadband maps and their global properties (e.g., surface brightness and absorption features). In Section \ref{sec:compare} we describe the comparison between the flux observed by \erosita\ and \rosat\ in detail. Section \ref{sec:features} discusses the appearance of some of the large spatial scale features in the maps and their differences.  Our conclusions are summarized in Section \ref{sec:summary}.

\section{Data processing}\label{sec:process}

\subsection{Calibration and processing logic}\label{sec:expmap}

The calibrated eRASS1 data were processed with the standard \erosita\ Science Analysis Software System pipeline (eSASS, version\_201009, developed by the German \erosita{} consortium \citealt{Brunner2022AA}). The eSASS pipeline is made up of task chains that aim to produce different data products \citep{Predehl2021AA}. The EXP chain we used includes procedures of packing the telemetry data into staging areas, quality good time interval (GTI) creation, telescope-merged event files and images creation, bright pixel or track flagging, and exclusion.  

In the eSASS pipeline calibration, the survey data are packed into distinct sky tiles according to the equatorial coordinates. Each sky tile is approximately 3.6 x 3.6 deg${}^2$ in size, with overlaps from the adjacent tiles. 

The maps presented in this paper were generated by assembling the exposures and count images of each sky tile, which were then projected onto the full sky while preserving the surface brightness. The primary procedure was carried out by Python and eSASS tools. We especially thank J. Sanders for the contribution of the projection script. The maps in HealPix format were derived with \texttt{healpy} package\footnote{http://healpix.sf.net}, \citep{HealpyZonca2019,Healpy2005ApJ}.

The creation of the surface brightness maps involves the following products and definitions: 
The counts map is the count of photons detected in the two-dimensional sky grids (i.e., per pixel: 9 arcmin$^2$), and the exposure map is the representation of the observing time distribution on the sky. The exposure is defined as the sampling of the on-axis exposure time summed over the used telescope modules after correction for the vignetting function (see details in Sec.~\ref{sec:vignetting}). The rates maps are the exposure-corrected maps obtained as the ratio of the counts map and the exposure map, and they are plotted in units of $\rm cts\, s^{-1}$. 

Both the counts map and the exposure map were derived from the same event-files. The event files are the tables including all the information on the counts detected in a sky tile (i.e., energy, detection pattern, and arrival time) The prefiltering applied on the event files was shared with the different products. We note that we only included data from the five telescope modules (TMs) equipped with on-chip filters (i.e., TM1, 2, 3, 4, and 6), and we excluded data from TM 5 and 7 because they were affected by a light leak  \citep{Predehl2021AA}.

\subsection{Exposure and vignette correction} \label{sec:vignetting}

\begin{figure}[t]
    \centering
    \includegraphics[width=0.5\textwidth, trim=13 49 -5 45,clip]{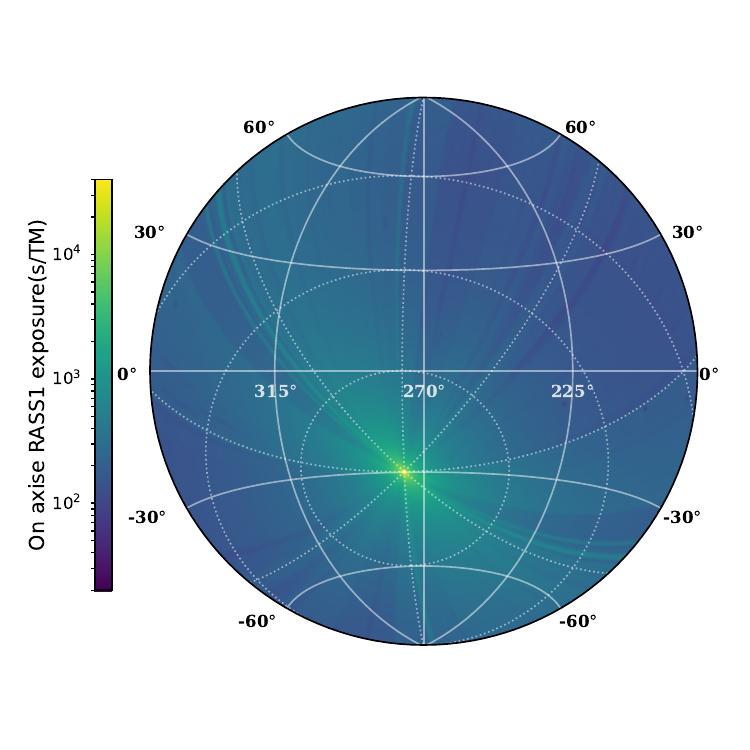}
    \caption{\small{eRASS1 unvignetted exposure in Galactic coordinates (solid grid) and ecliptic coordinates (dotted grid). Only the hemisphere (359.94423568 > $l$ > 179.94423568 degrees) accessible to \erosita{} DE consortium is presented. The color-coding corresponds to the accumulated integration time. The exposure was computed on-axis and averaged over the TMs (1, 2, 3, 4, and 6).}} 
    \label{fig:expo_example}
\end{figure}
The exposure image was produced by the eSASS task \texttt{expmap}\footnote{input parameters withvignetting=YES, withmergedmaps=YES, withweights=NO, withdetmaps=YES}. This setting produced the exposure corrected for the vignette and scaled to a single TM, as presented in Fig.\ref{fig:expo_example}. The correction factor for five TMs was implemented later in the effective area. \erosita{} scans the sky along the great circles that pass the ecliptic poles and has a precession rate of 0.167 deg $\rm {scan^{-1}}$ \citep{Predehl2021AA}, with a field of view with a diameter of 1.03 deg. As a result of the survey strategy, the \erosita{} exposure is characterized by a significant gradient across the sky. The cumulative on-axis observing time increases toward the ecliptic pole and reaches a maximum of $\sim$ 40 ks/TM (20 ks/TM as a vignetted exposure in 0.2--0.6 keV), while it is only $\sim 160$ s/TM (90 s/TM as vignetted exposure) close to the ecliptic plane. The overall averaged and vignette-corrected exposure of eRASS1 in the 0.2--0.6 keV band in the western hemisphere is about $ 170$ s/TM. A few small-scale fluctuations present in the exposure maps are caused by particular and time-dependent instrumental effects, such as TM switch off/on due to susceptibility in the camera electronics, variations in chopper settings, and so on.

\subsection{Instrumental background} \label{sec:instrbck}

Accurate modeling and subtraction of the instrumental background is key to retrieving low surface brightness features in the maps. The instrumental background in the X-ray CCD detectors mostly results from electronic readout noise and charges caused by cosmic rays hitting the detector. Regular monitoring of the particle background has revealed that it is largely featureless (i.e., it can be modeled by smooth functions), it increases toward lower energies, and it flattens out at 2 keV. At higher energies ($\geq5$ keV), the instrumental background dominates the incident flux in $\sim$ 90\% of the sky. Over the time period covered by eRASS1, this signal can be considered as time invariant \citep{Freyberg2020SPIE}.

Filter-wheel-closed data (FWC) are data taken with the filter wheel placed in closed position, preventing the camera from observing photons from the sky. The FWC model\footnote{Here the FWC of version c020\_v1.0 was used. More details can be found at the \erosita{} Early data release website: \url{https://erosita.mpe.mpg.de/edr/eROSITAObservations/EDRFWC}} was used to estimate the instrumental background. \erosita{} has performed several FWC observations during the performance verification (PV) phase and the survey at intervals. The PV FWC data have been published along with the \erosita{} Early Data Release. 
 
The instrumental background is unprocessed by the telescope optics and thus remains unaffected by the vignette. The instrument background can be characterized by the following formalism:
\begin{equation}
{\centering
    R_{\rm sky}=\frac{C_{\rm total}-R_{\rm FWC}\times T_{\rm nonvig}}{T_{\rm vig}}=\frac{C_{\rm total}}{T_{\rm vig}}-R_{\rm FWC}\times \frac{T_{\rm nonvig}}{T_{\rm vig}},}
\end{equation}
where $R_{\rm sky}$ is the total count rate from X-ray sources (both diffuse and point-like), $C_{\rm total}$ is the total number of detected events, $T_{\rm nonvig}$ and $T_{\rm vig}$ are the unvignetted and vignette-corrected exposures, respectively, and $R_{\rm FWC}$ is the instrumental background, which is built from the TM1, 2, 3, 4, and 6 averaged FWC model. We also refer to the $\frac{T_{\rm nonvig}}{T_{\rm vig}}$ ratio as the vignette coefficient.
 
We note that the characterization of the FWC data is under constant development and improvement by the \erosita{} team. New background particle models will potentially improve in the time and CCD-temperature dependence of the FWC data.

\subsection{Point source subtraction}\label{sec:subtract}

To reveal the diffuse emission, the bright point sources need to be subtracted.
We used the eRASS1 source catalog \citep{Merloni23}, which contains about 850,000 point-source candidates. The uneven depth of the survey exposure naturally leads to nonuniform detection limits. To avoid the selection bias, only the sources above a rates threshold were subtracted in the diffuse maps. 
We defined the threshold on the count rates, which applies to two different purposes:
In order to compare \erosita{} maps with those of \rosat{}, we retrieved the equivalent threshold from \rosat{} (Tab.2 in \citet{Snowden1997ApJ}) and converted it into \erosita{} observed rates, which corresponds to $\sim$0.30 $\rm cts\, s^{-1}$ in the 0.6--2.3 keV band.
A different threshold was instead chosen to include as many point sources as possible while retaining a uniform portion of the cosmic X-ray background from the entire sky.
We determined the latter threshold as approximately the count rate of the faintest source that can be detected in the low-exposure regions (i.e., close to the ecliptic plane) of 0.03 ct/s in the 0.2--5.0 keV band.

We created a cheese-like mask in which the pixels above threshold (i.e., including point sources) were masked out. For each source, a circular mask with a predetermined radius was applied. The radius was scaled according to the flux of the source. The minimum radius was $100 \arcsec$ (about six times the half-energy width of \erosita{} at 1.49 keV). The resulting masked regions were then filled with the average of the surrounding rates, estimated from the adjacent region where the signal to noise ratio (S/N) exceeds 2.

We note that all sources and features identified as extended sources in the eRASS1 catalog were preserved. Dust scattering can also produce a halo of diffuse emission around bright sources. This effect has been studied around different sources using the \rosat{} data \cite{Lamer2021AA,Jin2017MNRAS, Predehl1995AA}) and is clearly detected by \erosita{} as well (see Appendix D. Fig. \ref{fig:ero_r4}--\ref{fig:ero_r7}, the large halo surrounding Cir X-1, GX 349+2). We here retained these dust-scattering halos in the maps.

\section{Broadband \erosita\ diffuse emission maps}\label{sec:broadmap}

\begin{table*}
\centering
\begin{subtable}{\textwidth}
\centering
\begin{tabular}{cccc}
\hline
Energy Band & FWC Count Rate & FWC Flux & Vignette coef.\\
(keV)& ({\tiny$\rm ct~s^{-1}~deg^{-2}$})&  ({\tiny $\rm 10^{-7}~erg~s^{-1}~deg^{-2}~cm^{-2}~keV^{-1}$)}& \\
(1) & (2) & (3) & (4) \\
\hline
R45 (0.44-1.21) & 1.720 & 0.0170 & 0.558\\
R67 (0.73-2.04) & 2.810 & 0.0269 & 0.541\\
0.2--0.4 & 0.659 & 0.0095 & 0.543\\
0.4--0.6 & 0.527 & 0.0119 & 0.550\\
0.6--1.0 & 0.890 & 0.0166 & 0.562\\
1.0--2.3 & 2.713 & 0.0311 & 0.526\\
2.3--5.0 & 6.750 & 0.0600 & 0.413\\
\hline
\end{tabular}
\end{subtable}
\bigskip

\begin{subtable}{\textwidth}
\centering
\begin{tabular}{ccccc}
\hline
Energy Band  & $\frac{5}{7}$ECF${}_{5e20}$ & Count Rate${}_{\rm avg.}$  & Surface Brightness${}_{\rm avg.}$ & Total Flux${}_{\rm est.}$ \\
 ({\small keV}) & {\tiny ($\rm 10^{11}~ct~cm^{-2}erg^{-1}$)} & {\tiny ($\rm ct~s^{-1}~deg^{-2}$)} & {\tiny ($\tiny \rm 10^{-11}~erg~s^{-1}~deg^{-2}~cm^{-2}~keV^{-1}$)} & {\tiny $~(10^{-7}~\rm erg~s^{-1}~cm^{-2}~keV^{-1}$)}\\
 & (5) & (6) & (7) & (8) \\
\hline
R45 (0.44-1.21) & 9.51 & 10.75 $\pm$ 0.51 & 1.47 $\pm$ 0.07 & 3.03 $\pm$ 0.14 \\
R67 (0.73-2.04) & 8.13 & 7.40 $\pm$ 0.45 & 0.69 $\pm$ 0.04 & 1.43 $\pm$ 0.08\\
0.2--0.4 & 6.06 & 1.79 $\pm$ 0.14 & 1.45 $\pm$ 0.09& 2.99 $\pm$ 0.18\\
0.4--0.6 & 8.69 & 3.20 $\pm$ 0.17 & 1.85 $\pm$ 0.08 & 3.82 $\pm$ 0.16\\
0.6--1.0 & 9.84 & 4.85 $\pm$ 0.20 & 1.24 $\pm$ 0.03 & 2.57 $\pm$ 0.06\\
1.0--2.3 & 6.51 & 3.83 $\pm$ 0.22 & 0.46 $\pm$ 0.02 & 0.94 $\pm$ 0.04\\
2.3--5.0 & 0.82 & 1.02 $\pm$ 0.25  & 0.55 $\pm$ 0.09 & 1.17 $\pm$ 0.19\\

\hline
\end{tabular}
\end{subtable}
\caption{Global properties of the eRASS1 broadband map of the diffuse emission. All numbers are computed for the western sky ($180<l<360,~-90>b<90$). 
Columns: (1) Energy band. (2) The FWC count rate (i.e., instrumental background). (3) FWC flux. (4) Vignette coefficient. (5) ECF in units of $10^{11}$ cts\, cm${}^2$/erg. The fraction $\frac{5}{7}$ accounts for the number of TMs used. The spectral model assumed here is a power law with an index $\Gamma=2$ with absorption $\rm N_{\rm H}=2 \times 10^{20}~cm^{-2}$. (6) Image count rate averaged after subtracting the instrumental background. (7) Estimated surface brightness after subtracting the instrumental background. (8) Estimated total flux of the western sky ($180<l<360$).  The factors quoted here are based on the modeling assumptions and should be interpreted with caution due to the associated uncertainty. The errors quoted above are the statistical uncertainties.}\label{tab:global_pro}
\end{table*}
We present the first X-ray diffuse emission maps of eRASS1, available in Appendix E:Fig. \ref{fig:0.2_0.4}-\ref{fig:2.3_5.0}. The maps are represented in cylindrical projection, Galactic coordinates, in the energy bands: 0.2--0.4, 0.4--0.6, 0.6--1.0, 1.0--2.3, and 2.3--5.0 (keV). A spatially uniform contribution (in count rate units) of the instrumental background was subtracted from the maps, as described in Sec.\ref{sec:instrbck}. Adaptive smoothing (using S/N $\ge$ 20, unless otherwise stated) was applied for displaying purposes, while the unsmoothed version of the maps was used to quantify the properties reported in  Tab.\ref{tab:global_pro} and throughout the paper.
The pixel resolution was set to 9 arcmin$^2$ in all maps. This allows an average signal-to-noise ratio S/N$>1$ in the 0.2--1.0 keV band maps (see Fig.\ref{fig:significance}).

\begin{figure*}[]
    \includegraphics[width=0.95\textwidth,trim=52 70 57 70, clip]{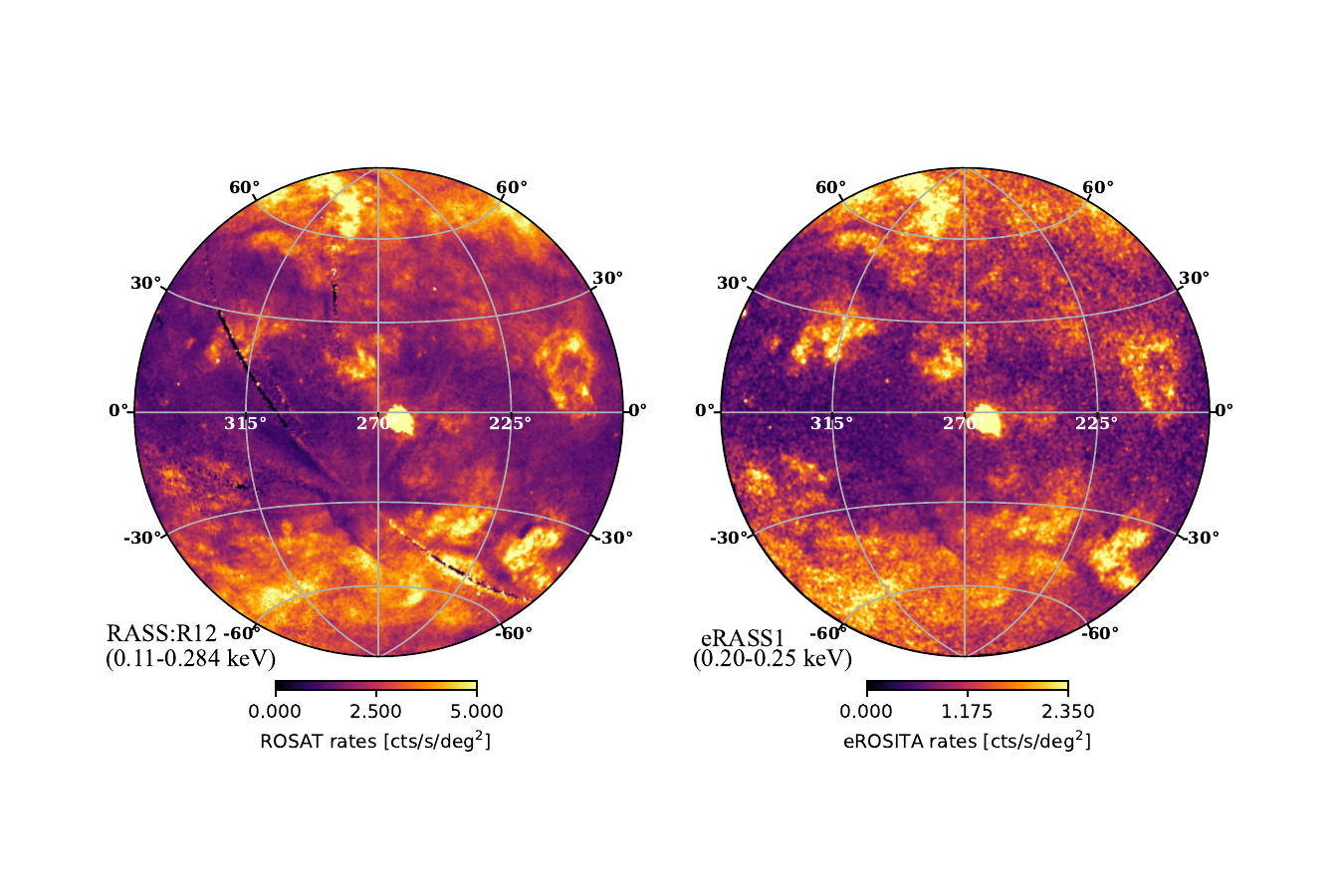}
    \centering
    \caption{Very soft band maps of \rosat{} (R12: 0.11--0.284 keV, left) and \erosita{}/eRASS1 (0.20--0.25 keV, right) in ZEA projection. The similarities between two surveys support the detectability of LHB with \erosita{}.}
    \label{fig:softband}
\end{figure*}
\begin{figure}[h!]
    \centering
    \includegraphics[width=0.5\textwidth,trim=25 0 -15 0, clip]{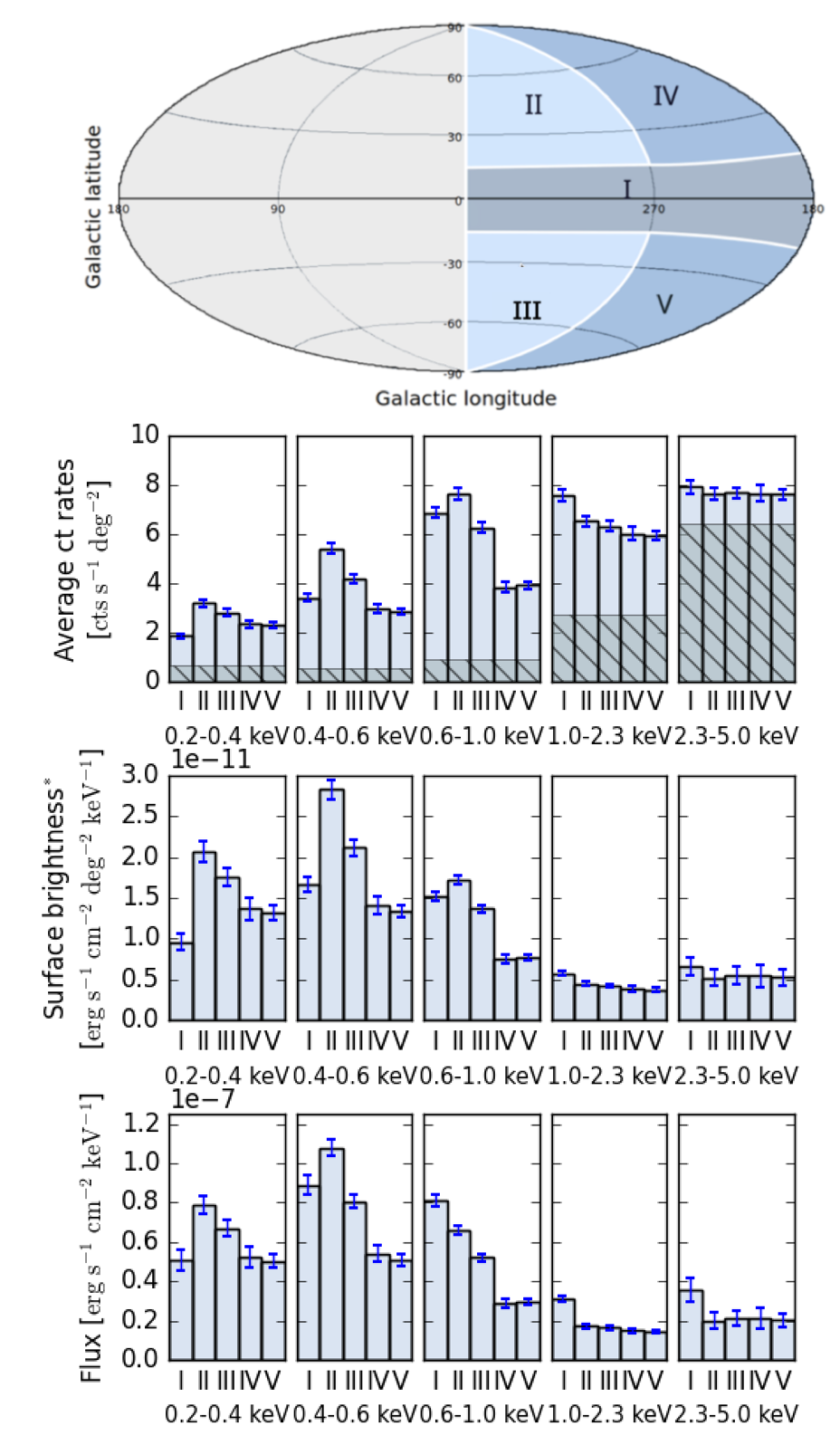} 
    \caption{Configuration and observed properties of the sky patches. We divided the \erosita{} DE sky into five parts as listed in Tab.\ref{tab:regions}. The hatched gray regions in the count-rate histogram show the level of the particle background. We converted count rates into surface brightness assuming a power-law model $S\propto E^{-\Gamma}$ with $\Gamma=2$ and absorption as in Tab.\ref{tab:regions}. The surface brightness and flux estimations use the ECFs reported in Tab.\ref{tab:global_pro}.}
    \label{fig:flux_est}
\end{figure}

At different energies (i.e., different bands), the relative importance of the difference components that build the diffuse emission changes, as shown by the spectral study of the \erosita{} Final Equatorial Depth Survey (eFEDS) region \cite{Ponti2023AA1}.

In the softest band (0.2--0.4 keV), the entire hemisphere shows a significant contribution from the LHB, modeled as an unabsorbed collisionally ionized plasma at $kT\sim0.1$ keV \citep{Snowden1998ApJ}. This emission must be situated within the local bubble because nonvanishing foreground emission toward the densest portions of the gas dark clouds is observed \citep{2023A&A...676A...3Y}). The morphology of the LHB was investigated using the \rosat{} data \cite{Liu2017ApJ}. For comparison, we present in Fig \ref{fig:softband} the softest band (0.20--0.25 keV) map of the eRASS1 data. The high degree of similarity between the \erosita{} and \rosat{} maps implies that \erosita{} has sufficient sensitivity to detect the very soft emission of the local plasma in the LHB.

A main emission source in 0.5--1.5 keV is the hot phase of the CGM of the Milky Way. This component has been fist recognized and detected by separation of the foreground and CXB components also through absorption by neutral clouds \citep{Snowden2000ApJS, Pietz1998AA}.
In a similar way, this component is also detected by \erosita{} and was studied \cite{2023arXiv231010715L} using the narrow-band data presented in a follow-up work of the present one (Zheng et al. in prep.);
In the same band, the diffuse X-ray emission carries potential information on the heliospheric SWCX. Because of the time variability, this component can be isolated and studied by \erosita{} (Dennerl et al. in prep). 

Above 0.7 keV, the emission from the CXB becomes the dominant component.
At higher energies ($E\gg1.5$~keV), the integrated emission from faint and unresolved X-ray point sources \cite[CXB,][]{Gilli2007AA, Brant2021arXiv211101156B} dominates the emission. In addition, a minor but non-negligible contribution may come from patches of very hot interstellar medium (ISM; $kT\simeq 0.7-1$ keV; \citealt{Ponti2023AA1}).

\subsection{Global estimate of the surface brightness and the flux}

In Tab.\ref{tab:global_pro} we list the observed properties (surface brightness and total flux) of the diffuse emission detected in the broadband maps. We report the estimates for the western hemisphere (359.94423568 > $l$ > 179.94423568) in the energy bands 0.2--0.4 keV, 0.4--0.6 keV, 0.6--1.0 keV, 1.0--2.3 keV, and 2.3--5.0 keV.
The ECF were estimated and used to convert the observed count rates into surface brightness ($S$). The computation of the ECF involves knowledge of the \erosita{} effective area and has to assume a spectral model. This was fixed as a power law with photon index of $\Gamma$=2.0 absorbed by a column density of $\rm N_{\ion{H}{I}}=5 \times 10^{20}~cm^{-2}$. The $N_H$ value stands for the median $N_H$ of of the HI4PI survey \cite{HI4PI2016AA}. The computed ECF factors are also listed in Tab.\ref{tab:global_pro}.

We note that in the computation of all the properties reported in Tab.\ref{tab:global_pro}, we excluded well-known bright extended sources such as Vela supernova remnant (SNR), Sco X-1, and the Virgo cluster, which are bright enough to otherwise contaminate the result. On these sources, we applied circular masks with radii of 5, 4, and 3 $\rm deg$, respectively.

\begin{table}[htbp]
    \centering
    \small
    \caption{Regions and areas}
    \begin{tabular}{c@{}c@{}c@{}c@{}c}
        \hline
        Region  & $l$ & $b$ & Area & ${N}_{\ion{H}{I}}^{\rm median}$  \\
        Denotation & (deg) &(deg) & ($\mathrm{deg}^{2}$) & ~~({\tiny$10^{20}$} cm${}^{-2}$)\\
        \hline
        I {\tiny(Disc)} & ~~[180, 360] & ~~[-15, +15] & ~~~5338.5 & 21\\
        II {\tiny(Galactic outflow N)} &~~[270, 360] & ~~[+15, +90] & ~~~3822.0 & 5.0\\
        III {\tiny(Galactic outflow S)} & ~~[270, 360] & ~~[-90, -15] & ~~~3822.0 & 3.7\\
        IV {\tiny(Galactic anti-center N)} & ~~[180, 270] & ~~[+15, +90] & ~~~3822.0 & 3.2\\
        V {\tiny(Galactic anti-center S)} & ~~[180, 270]& ~~[-90, -15] & ~~~3822.0 & 3.0\\
        \hline
    \end{tabular}\label{tab:regions}
\end{table}
\begin{figure}[b!]
    \centering
    \includegraphics[width=0.45\textwidth, trim=10 0 10 10,clip]{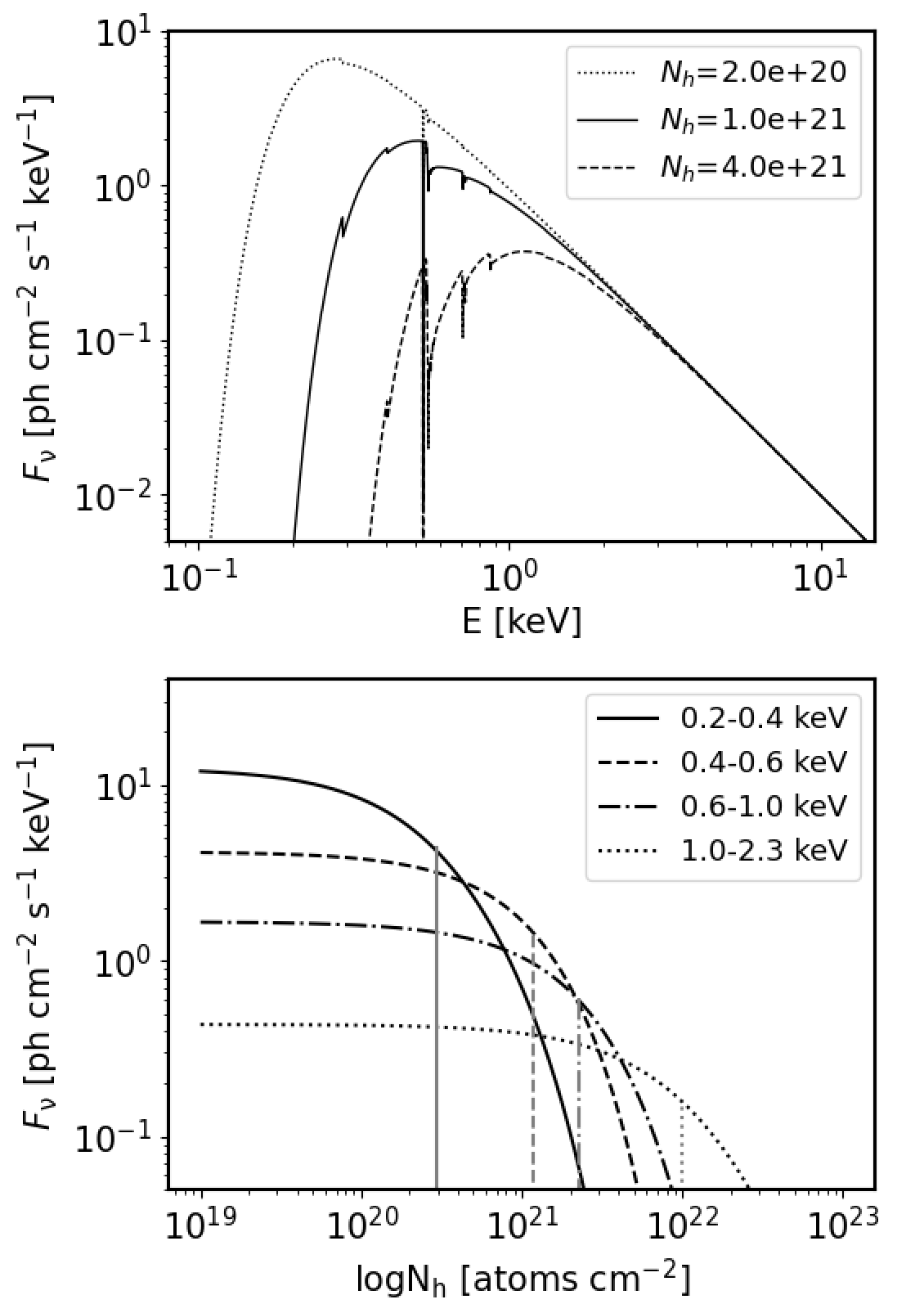}
    \caption{Effect of absorption on the X-ray spectrum. Top panel: Energy dependence of the absorption spectrum. The unabsorbed spectrum is defined as $F_{\nu}$=$\nu^{-2}$. The labeled lines show column densities matching the contours in Fig.{\ref{fig:HI4PI_absorption}}.
    Bottom panel: Effect of $N_{\ion{H}{I}}$ on the broadband flux. The vertical lines show the threshold at which $F_\nu$ decreases by a factor $\frac{1}{e}$ of the unabsorbed value in the same band.}
    \label{fig:Absorption}
\end{figure}
\begin{figure*}[]
    \includegraphics[width=0.95\textwidth,trim=31 17 13 19, clip]{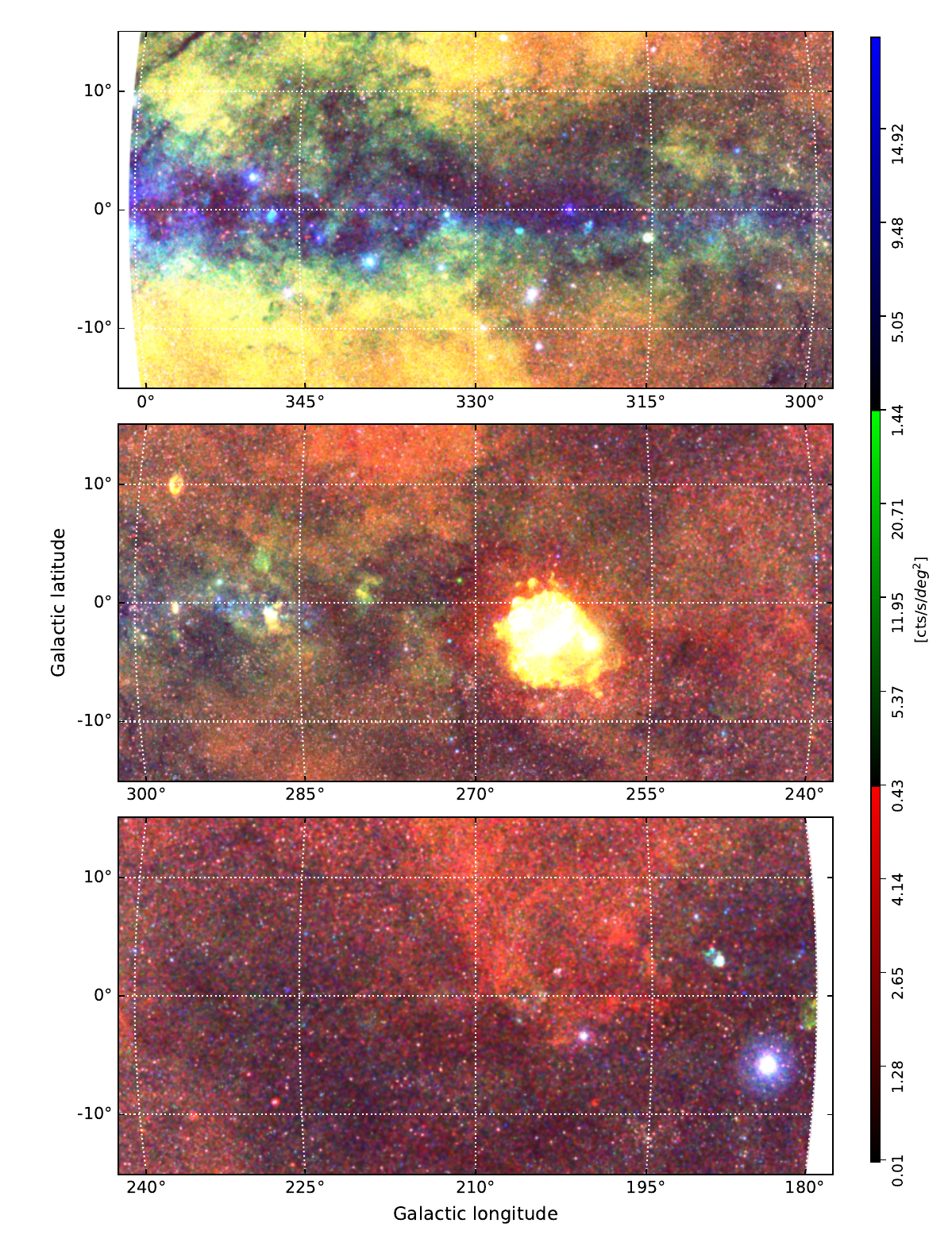}
    \centering
    \caption{Maps of the Galactic disk from eRASS1 survey in RGB colors (Red: 0.2--0.5 keV. Green: 0.5--1.0 keV. Blue: 1.0--2.0 keV). The color brightness scales as $\rm S_{R,G,B }=\frac{log(1000*x+1)}{log(1000)}$. The maps have a resolution of 0.25 arcmin$^2$ and are adaptively smoothed using $\rm S/N_{sm}\equiv 5$. The orthographic projection (SIN) is used. The point sources present in the image have been kept.}
    \label{fig:disc}
\end{figure*}
\begin{figure*}[t!]
    \centering
    \includegraphics[width=0.95\textwidth, trim=40 0 0 0,clip]{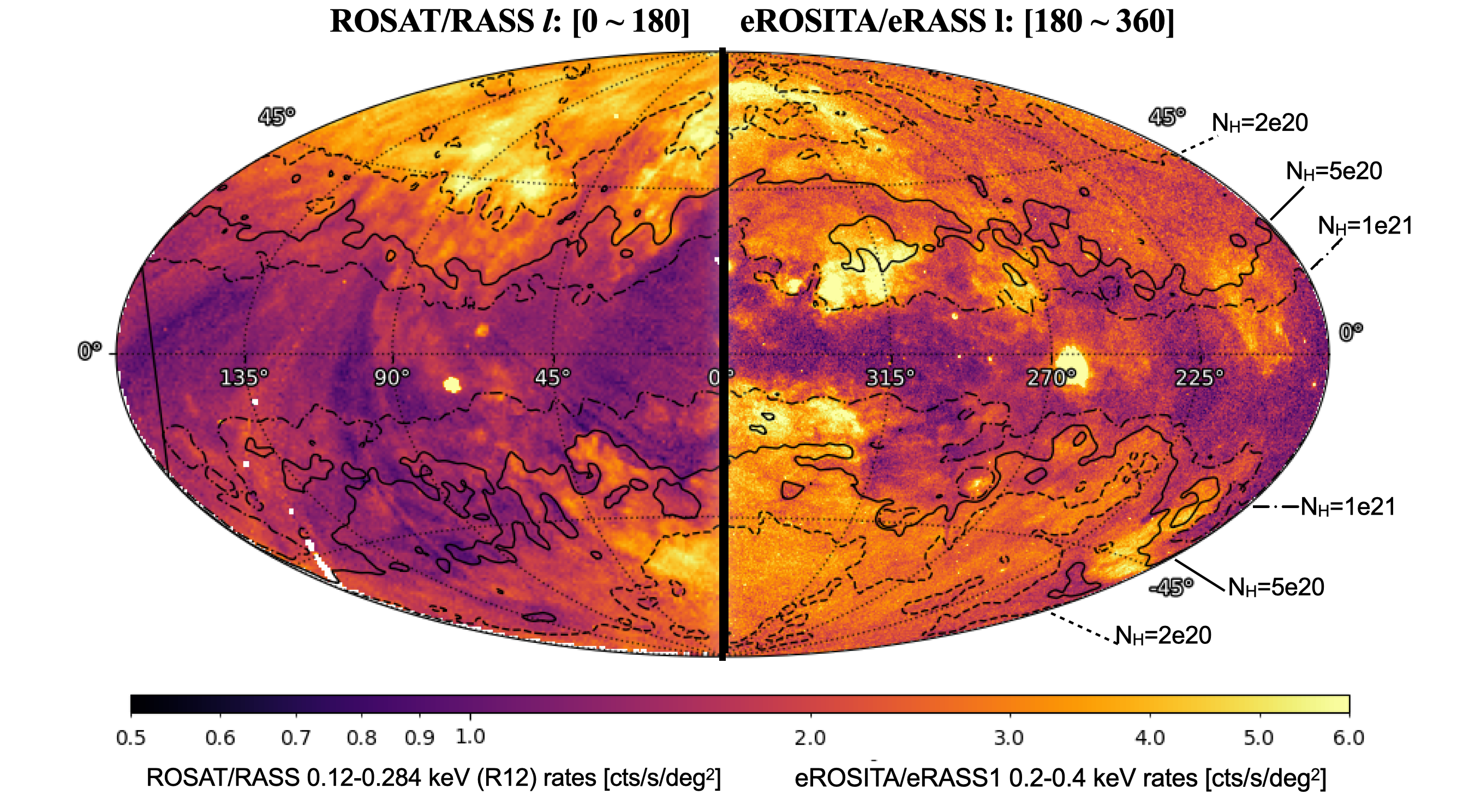}
    \caption{ $N_{\ion{H}{I}}$ contours overlaid onto the \erosita{} soft-band images. The western sky $l$=[180, 360] shows the \erosita{}/eRASS1 data in the 0.2--0.4 keV band mirrored with respect to Galactic longitude $l=0$. 
    The eastern sky $l$=[0, 180] shows the \rosat{}/RASS R12 (0.11-0.284 keV) band map. The contour superimposed are the column density from the HI4PI survey: $\rm N_{\ion{H}{I}}=2\times10^{20}$ (dotted), $5\times10^{20}$ (solid), and $1\times10^{21}~\rm cm^{-2}$ (dot dashed).}
    \label{fig:HI4PI_absorption} 
\end{figure*}

We divided the western sky into five distinct patches I-V, based on quadrants, variations in absorption, and the predominant source of diffuse emissions (see Tab.\ref{tab:regions}):
Region I is representative of the Galactic plane, where the disk absorption has a significant impact. Region I spans the range of Galactic latitudes -15 to 15, roughly matching the $\rm N_{\ion{H}{I}}=10^{21} cm^{-2}$ contour of $N_H$ \citep{HI4PI2016AA}. 
Regions II and III represent the area of the northern and southern Galactic outflow (i.e., the \erosita{} and Fermi bubbles), respectively. 
Regions IV and V are mostly free from either outflows or foregrounds. They sit in the northern and southern hemisphere, respectively, and lie closer to the Galactic anticenter. The absorption in regions IV and V has a smaller impact than for the rest of the western sky.

The averaged count rates, the surface brightness, and the total flux of regions I--V are presented in Fig.\ref{fig:flux_est}. In general, the large-scale surface brightness and flux peak in the 0.4--0.6 keV band. Below 0.4 keV, absorption due to the Galactic disk neutral material significantly reduces the observed flux. The uncertainties were derived as the statistical error.

We note that region II consistently shows the highest surface brightness of the regions up to 1 keV. The contrast is particularly noticeable in the 0.4--0.6 keV band, where region II exhibits a surface brightness of 2.75 $\times 10^{-11} \rm erg/s/{cm}^2/deg^2/keV$.
An enhancement toward the disk is well noticeable across the sky, especially in the harder bands, such as 2.3--5.0 keV.
Most regions exhibit their highest surface brightness in the 0.4--0.6 keV band, while region I peaks in the 0.6--1.0 keV band, likely because of the heavy absorption as well as the contribution from bright point sources and the ISM in the Galactic plane.

\subsection{Absorption features} \label{sec:absorption}

The observed soft-band diffuse X-ray maps clearly exhibit absorption as a significant suppression of the emission. In Fig. \ref{fig:Absorption} we present examples of the effect of absorption on the X-ray spectrum. The decrement was modeled as an exponential with an exponent proportional to the column density $N_H$ of the absorbing material and an energy-dependent cross section $\sigma \propto {\mathrm E}^{-3}$ \citep{ Balucinska1992ApJ, Wilms2000ApJ, Locatelli2022AA}.

The main contribution to the absorption comes from the Galactic interstellar medium. The cold phases encompassed by the ISM, including dust and molecules (dominantly H$_2$) and mostly traced by neutral hydrogen, are in fact mainly concentrated in the Galactic plane \citep{HI4PI2016AA}, causing the most severe absorption of the X-ray intensity in region I. Toward the high galactic latitude sky, the photoelectric absorption becomes less important as the overall density distribution scales as sec|b| \citep{Koutroumpa2006AA, Lallement2016AA}. In addition, individual local clouds casts their X-ray shadows, as already probed by \rosat{}\citep{Snowden1991Sci, Snowden2000ApJS}, and \erosita{} \cite{2023A&A...676A...3Y}.

Fig \ref{fig:disc} shows the X-ray emission along the Galactic plane from eRASS1 in RGB colors (Red: 0.2--0.5 keV. Green: 0.5--1.0 keV. Blue: 1.0--2.0 keV). The color-code is log-scaled and computed as $\rm S_{R,G,B }=\frac{log(1000*x+1)}{log(1000)}$. We note that the point sources in Fig.~\ref{fig:disc} have been kept. 

The dark band along the Galactic plane caused by absorption is one of the most noticeable features in \erosita{}/eRASS1 maps. The most severe Galactic absorption is observed close to the direction of the Galactic center, where blue points to the hard-band photons, which alone pierce the foreground absorbing screen. The top panel of Fig \ref{fig:disc} clearly shows the sharp dimming of the emission as a function of Galactic latitude. The X-ray radiation dominant at high latitudes therefore has to originate behind the high column density regions in order to be obscured and to produce the darker patterns visible in region I. The gas clouds at medium latitudes can thus serve as useful distance constraints to the diffuse emission at the largest angular scales.
To provide some reference values, by defining the on-plane absorption horizon as the distance from which the observed emission is reduced by a factor $1/e$, an observer at the Sun would see through the disk up to $\sim$0.1, 0.4, 0.8, and 3 kpc $\frac{\rm 1~cm^{-3}}{N_{\ion{H}{I}}}$ in band 0.2--0.4 keV, 0.4--0.6 keV, 0.6--1.0 keV, and 1.0--2.3 keV, respectively. For comparison, the distance from the Solar System to the Galactic center is approximately 8.1 kpc\citep{Reid2004ApJ}.

In addition, the top panel shows an enhanced 0.5--1.0 keV emission that outlines the base of the hot plasma of the \erosita{} bubbles. In the middle panel of Fig \ref{fig:disc}, the 0.2--0.5 keV emission and the 0.5--1.0 keV become relatively mixed and even, with the blue of 1.0--2.3 keV photons highlighting the hard point sources in the disk. In the direction of the Galactic anticenter (bottom panel of Fig \ref{fig:disc}), the emission is dominated by soft emission of 0.2--0.5 keV, especially in the region of bright SNRs. The Monogem Ring (Knies et al. in prep) is visible as an extended ($\sim 10$ deg) arc-shaped emission. The blue diffuse emission above 1.0 keV is barely noticeable, while the brightest peak corresponds to hard point sources.

\begin{figure*}[]
\includegraphics[width=0.95\textwidth,trim=420 0 450 0, clip, angle=0]{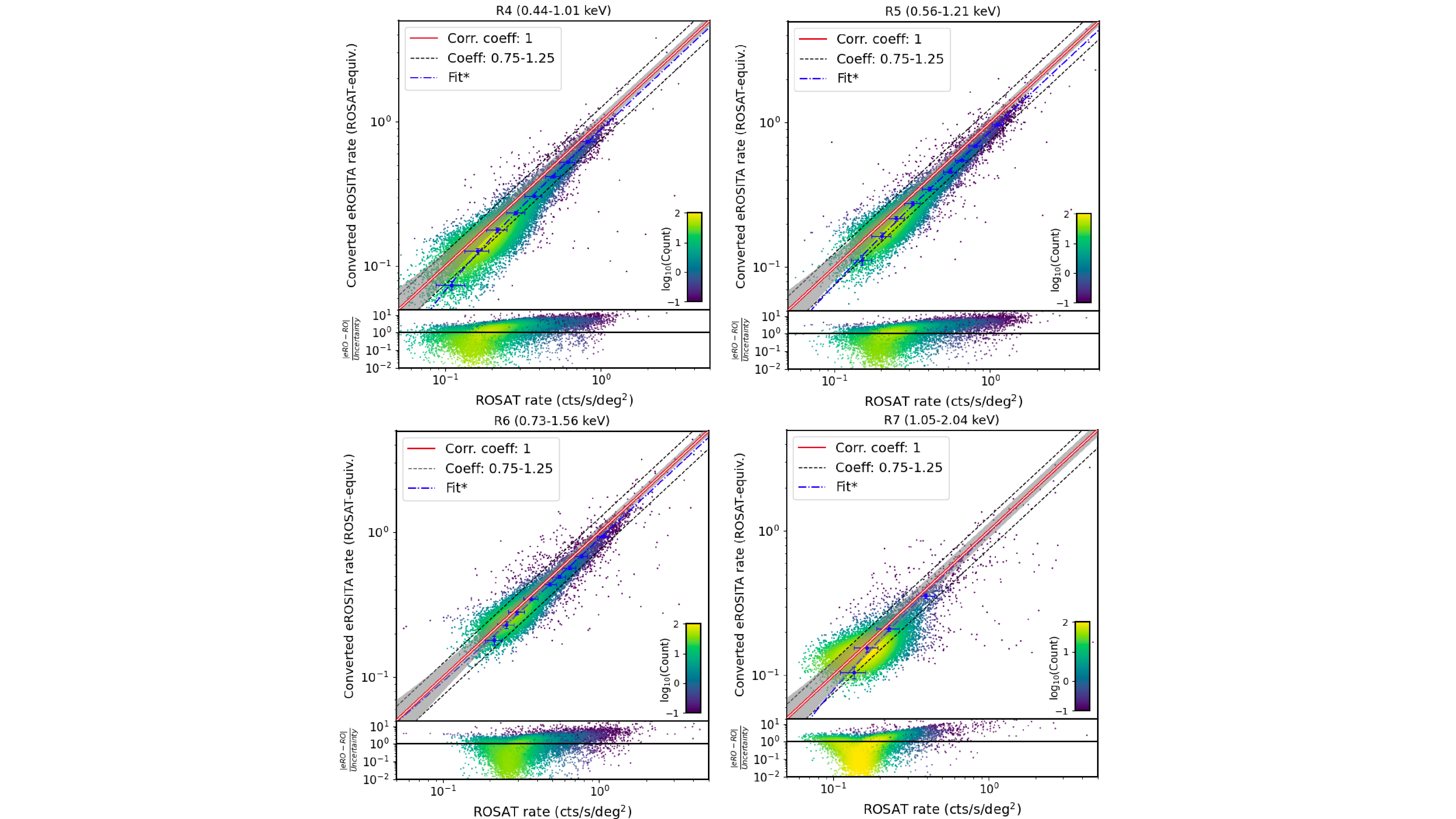}
    \caption{Rate comparison of diffuse emission maps of \erosita{}/eRASS1 and \rosat{}/RASS in log scale. 
    Each data point shows the rates from 0.83 $\rm deg^{2}$ sky region (NSIDE 64). The converted \erosita{} rates were calculated from Eq.\ref{eq:Rcalc} to be equivalent to the \rosat{} response. The screening is by $\rm S/N > 3$ in the \rosat{} data. The color corresponds the density of the data in this plot. 
    The navy points with error bars mark the density peak in the \erosita{} rates with a 1 $\sigma$ confidence.
    The navy dash-dotted line is the best fit of the densest points.
    The lines with a slope of 1 and 0.75 (1.25) and zero intercept are shown with the solid red line and dashed black lines, respectively.
    The intrinsic Poisson uncertainty level corresponding to 1 pix size is shown by the gray shadow.}
    \label{fig:4relations}
\end{figure*}

\begin{figure*}[htbp!]
\includegraphics[width=0.97\textwidth,trim=0 68 4 45, clip, angle=0]{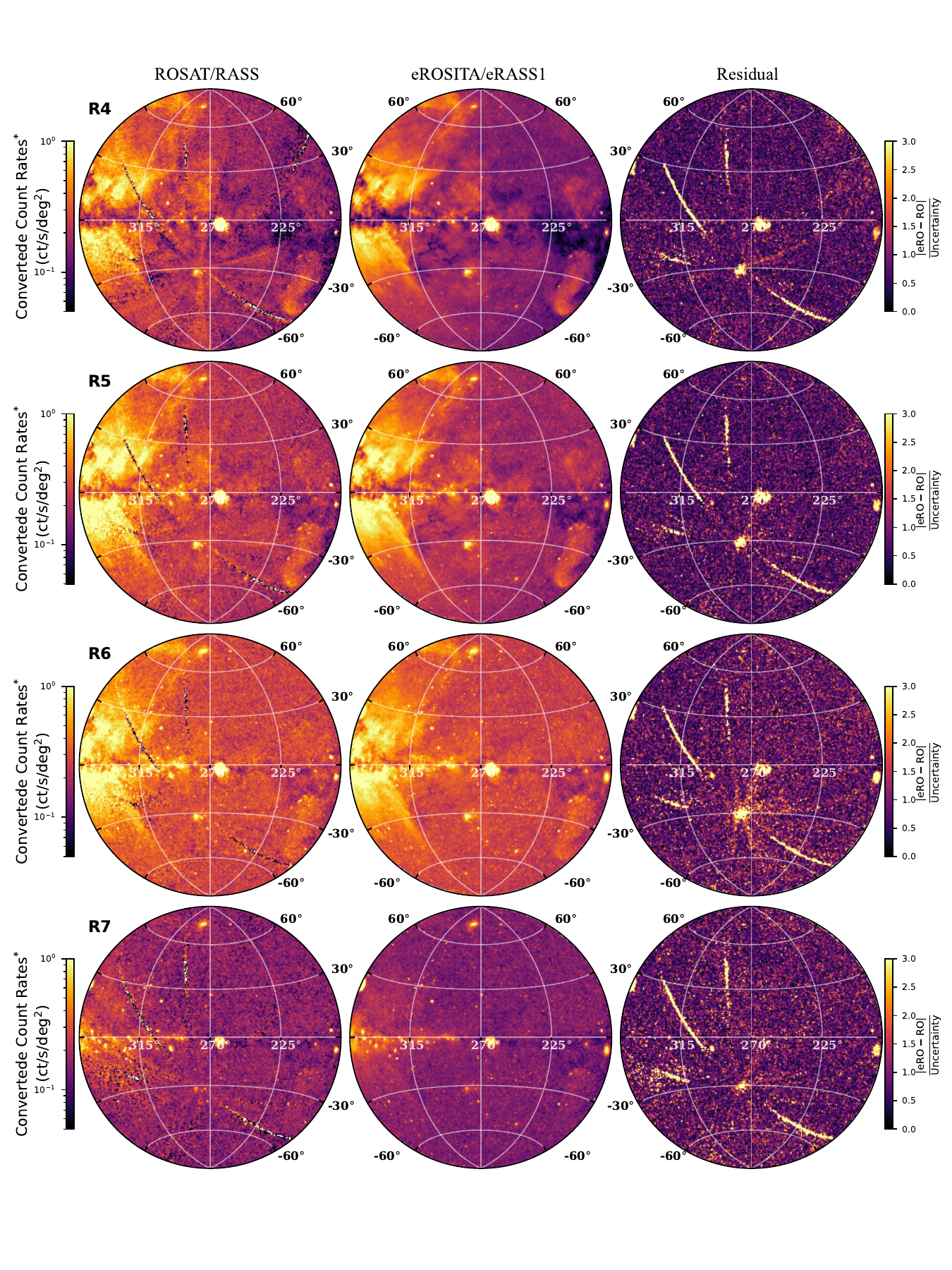}
    \caption{Visual comparison of the data. \rosat{}/RASS maps (left column), the converted \erosita{} maps (middle column), and the residual maps (right column) of the diffuse emission observed in four bands. The converted count rates are the rates scaled with the \rosat{} PSPC response. The deviation is defined as the $\frac{|eRO-RO|}{\rm Uncertainty}$. The applied ZEA projection is centered at l, b = 270.0, 0.0.}
    \label{fig:residual}
\end{figure*}

In Fig.\ref{fig:HI4PI_absorption} the contours from the column density {map taken} from the full-sky neutral hydrogen survey \citep{HI4PI2016AA} overlap the 0.2--0.4~keV eRASS1 map.
The RASS data are affected by systematic uncertainties that can be reached by the apparent stripes, while the eRASS data are nearly smooth in brightness transition.
We show three ${N_{\ion{H}{I}}}$ contour levels as the imprints of the Galactic absorption gradient: $2\times10^{20}$, $5\times10^{20}$, and $1\times10^{21}~\rm cm^{-2}$.
The anticorrelation of the column density and the X-ray emission is clear.
At energies above several kilo electronvolts, the emission from the Galactic disk is only marginally absorbed (see also Fig.\ref{fig:1.0_2.3} and Fig.\ref{fig:2.3_5.0}).

\begin{figure*}[]
    \includegraphics[width=0.95\textwidth,trim=280 0 340 0, clip]{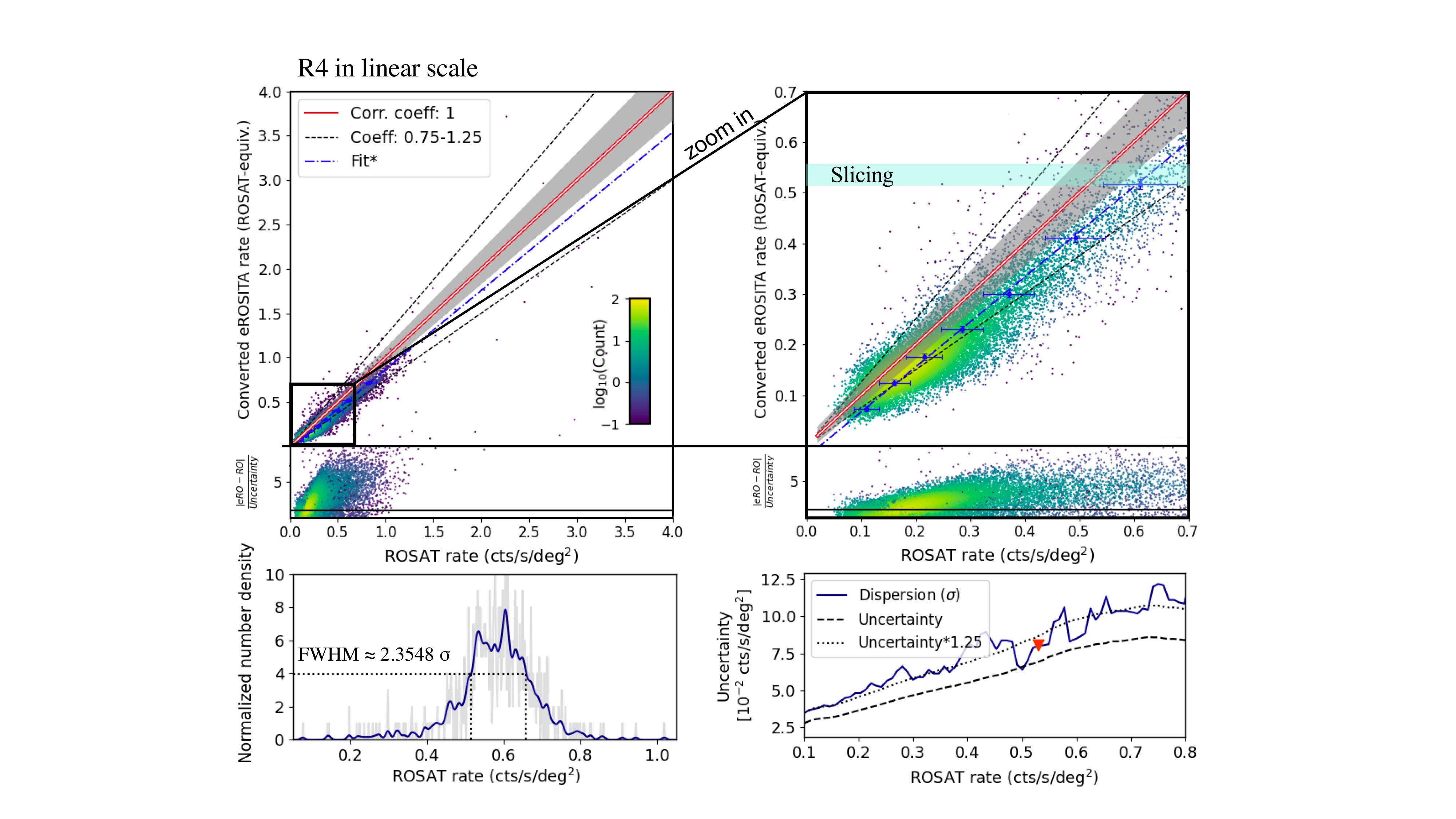}
    \caption{Detailed view of the \rosat{} and \erosita{} comparison in the R4 energy band.
    Top left: Same as Fig.\ref{fig:4relations} (top left panel), shown in linear scale.
    Top right: Zoom-in on the rate range $0-0.8 \rm cts\, s^{-1}\, deg^{-2}$.
    Bottom left: 
    Schematic distribution of the \rosat{} rates taken from the redefined \erosita{} rate range (shown in cyan in the top right panel). 
    The blue line is the smoothed number density that falls into this rate channel, from which the FWHM is measured.  
    Bottom right: Equivalent $\sigma$ and the uncertainty reported by \rosat{} with respect to the \erosita converted rates. 
    The \rosat{} uncertainty is dominated by Poissonian noise. 
    The red triangle shows the equivalent $\sigma$ from the bottom left panel.
    The 1.25 times \rosat{} uncertainty is shown with a dashed line for comparison.}
    \label{fig:uncertainty}
\end{figure*}

\begin{figure}[]
    \centering
    \includegraphics[width=0.475\textwidth, trim=10 45 5 30, clip]{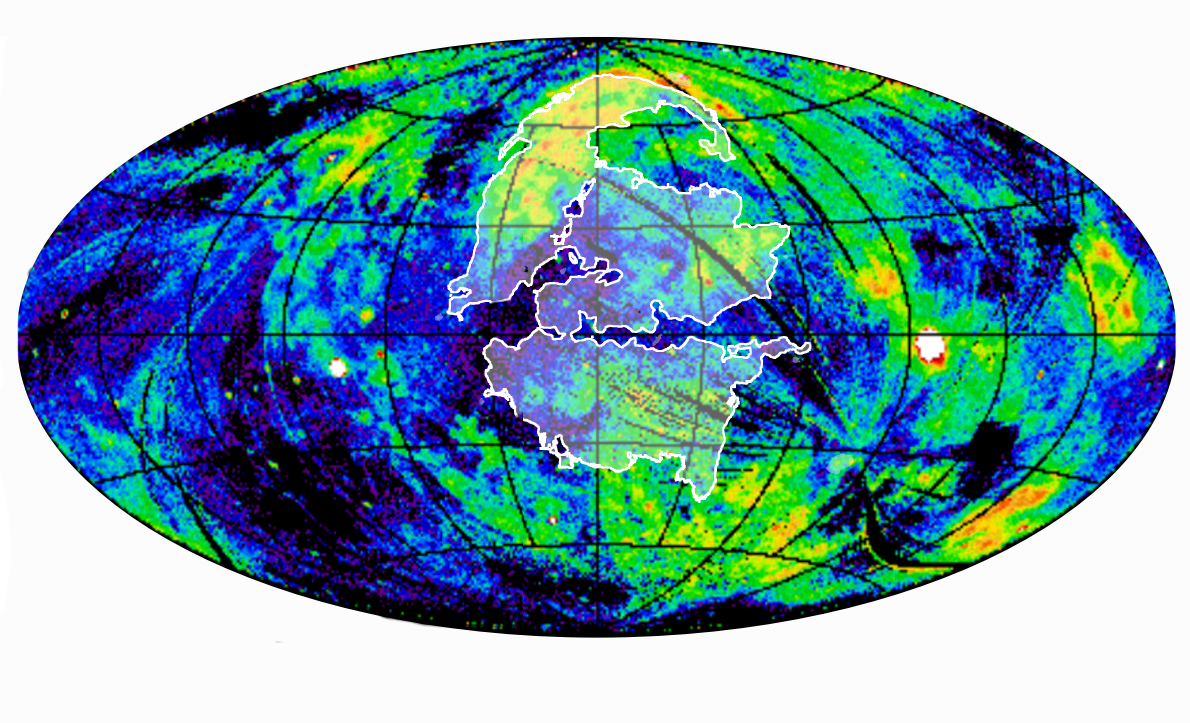}
    \caption{\small{Foreground components in the \rosat{} map with the eROSITA bubble contour overlaid. This foreground maps was built by \cite{Freyberg1994PhDT}, who modeled and separated the foreground emissions from RASS map. The resulting maps are shown with the angular resolution of 0.2\degree. We overlay the contour of eROSITA bubbles (defined from 0.6-1.0 keV band) as a white translucent area.}} 
    \label{fig:Michael2}
\end{figure}

\section{Comparison with \rosat{} X-ray diffuse emission} \label{sec:compare}

We compared the \erosita{} broadband maps to the \rosat{} diffuse background maps from \citep{Snowden1997ApJ}\footnote{available at: \url{https://heasarc.gsfc.nasa.gov/FTP/rosat/data/pspc/images/sxrb_maps/xray_highres_maps}, or \url{http://www.xray.mpe.mpg.de/rosat/survey/sxrb/12/fits.html}}. The \rosat{} maps were corrected for particle background and scattered solar X-rays along with the long-term enhancements (LTE) detected by \rosat{}. A few regions that were severely impacted by the effect of single reflection (i.e., the photons are only reflected once before they reach the sensor) were also manually removed in the version of the \rosat{} maps presented in this work.

The left and right columns in Fig. \ref{fig:ero_r4}--\ref{fig:ero_r7} present the count rates of the diffuse emission observed by \erosita{} and \rosat{}, respectively. The maps are projected with the zenith equal area (ZEA) projection and use Galactic coordinates in the \rosat{} R4, R5, R6, and R7 energy bands (Fig. \ref{fig:ero_r4} to \ref{fig:ero_r7}, respectively; see Tab.\ref{tab:ro_bands} and \citet{Snowden1997ApJ} for the definitions of the energy ranges). Identical sources yield different count rates in the \erosita\ and \rosat\ data due to their different effective areas (see Fig. \ref{fig:effective_area}). We employed energy-to-counts conversions in the relevant energy range with the assumption of a power law $\nu^{-2}$ spectrum to convert rates into fluxes. The color bars in Fig. \ref{fig:ero_r4}--\ref{fig:ero_r7} were set to the same flux values to ensure that features with constant flux appeared with similar colors in both maps. 

A remarkable match of the X-ray emissions detected by \erosita\ (eRASS1) and \rosat\ (RASS) is immediately noticeable. Prominent large-scale features exhibit comparable levels of brightness in both measurements. The sites showing the largest deviation between the maps form stripes along ecliptic longitudes in the \rosat{} maps. These are caused by the enhanced particle background, induced by solar activity, at the orbit/position of the \rosat\ spacecraft, compared to the relatively quiet environment of SRG/\erosita{} (i.e., the L2 point). Additional residual effects close to the passage of \rosat{} through the South Atlantic Anomaly are also present \citep{Snowden1993ApJ}.
Another obvious discrepancy is visible around Sco X-1 (i.e., the brightest point source in the X-ray sky), where the contribution due to the single reflection is brighter in the \erosita\ data.  With their large photon-collecting area, the \erosita{} eRASS1 maps are able to probe the diffuse emission more deeply. In Fig.\ref{fig:cluster}, a discrete but large number of extended sources (i.e., galaxy clusters) is now also recognizable in the \erosita{} maps, but they could not be distinguished in the \rosat{} maps.

\subsection{Method}

To assess the consistency of the \erosita\ and \rosat\ measurements, we compared the observed count rates. To do this, the maps were smoothed on a common grid to an angular resolution of 0.839 deg$^2$\footnote{this angle corresponds to an Healpix pixel using {\tt nside=64}. }. In order to avoid assuming the spectral shape of the emission, the comparison was carried out in count rates rather than flux. Different sources are characterized by different spectra. ECFs computed for each pixel, depending on the assumed spectrum, would be required to take this into account.

The observed count rates were determined by the numbers of photons within each pixel divided by the exposure (vignette corrected). To make the observed \erosita\ count rates directly comparable to the \rosat\ ones, we split each broad energy band (R2, R4, R5, R6, and R7) into ten narrower energy bands (this number of narrow bands still preserves S/N $>$ 3). For a sufficiently small energy range, the converted \erosita\ count rate becomes equivalent to that of \rosat\, regardless of the spectrum of the source. We calculated the \erosita\ count rate for each sub-band and multiplied it by the ratio of the effective areas of the two different instruments. Finally, we summed the contribution from all narrow energy bands to present the equivalent rate in the broad band. 
In this way, the observed \erosita\ count rate was converted into a \rosat\ equivalent count rate, becoming directly comparable to that of \rosat.

The procedure is formalized by equation~\ref{eq:Rcalc}, where $\rm E_{i}$ represents the differential energy bins in each band,

\begin{equation}
{\centering
R_{\rm ero2ro}=\sum_{\mathrm{E}_{\mathrm{i}}}\left(\frac{R_{\mathrm{eROSITA}}\left(\mathrm{E}_{\mathrm{i}}\right)}{\mathrm A_{\mathrm{eROSITA}}\left(\mathrm E_{i}\right)} * \mathrm A_{\mathrm{ROSAT}}\left(\mathrm E_{i}\right)\right).}
\label{eq:Rcalc}
\end{equation}

For completeness and reproducibility, we note that we applied the following corrections before computing the converted count rates for \rosat{} and \erosita{}: 
\begin{itemize}
    \item[i)]{Large-scale rebinning of \rosat{} data: the removal of particle background, scattered solar foreground, and long-term enhancement in the publicly available \rosat{} soft X-ray maps \citep{Snowden1997ApJ} leaves pixels with negative or unusually high count rates. These spurious pixels are mostly found along the ecliptic longitude. By rebinning the maps into larger pixels, the high-value pixels are naturally removed.}
    \item[ii)]{S/N screening. We used the uncertainty (sigma) map of \rosat{} as a quality criterion to screen out pixels with S/N$<3$, mostly tracing pixels affected by a systematic large uncertainty. For example, applying S/N $>3$ to the R4 map preserves 98.3\% of the data points, with 15.7\% of them holding S/N $>$ 10. The effect of filtering the data with higher S/N is more noticeable in band R7.}
    \item[iii)]{Single reflection: Due to the single reflection caused when \rosat{} and \erosita{} scan over very bright sources (e.g. Sco X-1), the observed flux can experience a systematic increase due to the photons coming from outside the field of view (about 1 deg in diameter). These spurious photons have interacted with the mirror system only once before ending up on the detector plane. The single reflection is stronger in \erosita{} than in \rosat{} because of the extra baffle of \erosita{} and the large reflection angle. This effect is visible around Sco X-1 and other bright point sources. We masked these sources with a $5\degree$ radius circle. As visible in Fig.\ref{fig:correction1} (c), the filtering on single reflections affects the typical count rates in the $0.5< R_{ro}< 1.0$ and $~0.8<R_{ero}<2.0$ of \rosat{} and \erosita{}, respectively.}
\end{itemize}

\subsection{Results}
The converted \erosita{} count rates (i.e., equivalent to \rosat; see eq.~\ref{eq:Rcalc}) are plotted against the \rosat{} rates in Fig.\ref{fig:4relations} for the R4, R5, R6, and R7 bands.
The density peak at fixed \erosita{} count rate is marked by a blue dot with the y(x) error bars showing the converted \erosita{} and \rosat{} uncertainty, respectively. The uncertainty in the \erosita{} maps is at least five times smaller than in the \rosat{}, at all rates. The rate (Poissonian) uncertainty of the \erosita{} and \rosat{} counts is also applied around the 1:1 relation in the shaded gray region. 

Fig.\ref{fig:4relations} shows and evident correlation between the two instruments in all bands in the $\rm  0.05-10\, cts\, s^{-1}\, deg^{-2}$ rate range. The 1:1 relation is generally well matched by the data, with a scatter in addition. As shown by the yellow glow in the plots, most of the data are in the $\rm  0.1-0.3\, cts\, s^{-1}\, deg^{-2}$ rate range. The converted rates of \erosita{} in R4 and R5 appear to be slightly lower than the \rosat{} rates in the entire range of values, pointing to a small systematic shift. In R6 and R7, the rates agree with the 1:1 relation, but they appear to be more scattered in R7. 
R7 in particular holds more data at high rates that are affected by large deviations from the 1:1 relation compared the other bands. For the higher photon energies of R7, the Galactic diffuse emission becomes fainter and the CXB contributes most at the faint end, which follows the 1:1 relation better.

The lower panels in Fig.\ref{fig:4relations} show the residuals of the data with respect to the 1:1 relation. The residuals are defined as $\rm |eRO-RO|\sigma_{tot}$, where $\rm \sigma_{tot}$ refers to the combined Poisson uncertainty of both \erosita{} and \rosat{}. The residuals averaged over the western sky are 1.079, 1.073, 1.424, and 1.240 in R4, R5, R6, and R7, respectively. Of the four bands, R6 thus shows the largest scatter. Data with lower rates are in general closer to the 1:1 relation, considering their larger uncertainty, while the bright end deviates more from the 1:1 line in all bands.

The blue points summarize the trend followed by the data. We fit these peak density points with a linear relation. The result is shown by the dash-dotted blue lines in all the panels of Fig.\ref{fig:4relations}. The best-fit relations between the converted \erosita{} rates (y) and the \rosat{} rates (x) are $0.905 x-0.022,\, 0.880x-0.01,\, 0.900x+0.003$ and $0.992x-0.020$ for R4, R5, R6, and R7, respectively. From these fits, we find that the converted \erosita{} rates are lower than the \rosat{} rates by about $\sim 12\%$, except for the R7 band. We note that we only fit the densest location instead of the whole data because a small number of extreme outliers had a very small statistical uncertainty because we did not include any systematic uncertainty in the computation. Thus, the statistical uncertainty underestimates the total uncertainty by an amount comparable to the (unknown) systematic, potentially arising from the removal of solar activity enhancement and subtraction of the particle background.

In Fig.\ref{fig:residual} (see Fig.\ref{fig:ero_r4}-\ref{fig:ero_r7} for zoom-in versions) we show the rate maps of \rosat{} ($R_{\rm ro}$, left column) and the converted \erosita{} ($R_{\rm ero2ro}$, central column) rate maps for direct visual inspection. After the conversion, these maps show theoretically comparable rate measurements. The places where \erosita{} is brighter than \rosat{} indicate the points above the 1:1 correlation found in Fig.\ref{fig:4relations} and vice versa. The coherent patches producing the highest residuals are clearly visible in the left and right column of Fig.~\ref{fig:residual}. The converted \erosita{} maps are visually similar to the \rosat{} rates, with most of the diffuse structures being similar. The residual maps show that the most prominent difference appears as bright stripes along ecliptic longitudes, indicating a dependence on the observing time. These stripes arise from the subtraction of the \rosat{} long-term enhancement and short flares, done empirically. The traces can be also seen directly from the cleaned \rosat{} maps in the left column. The dependence on the observation time can explain the difference at fainter count rates, and also the intercept at zeropoint of the fitted relations: 
The \erosita{}/eRASS1 was in fact carried out during a period of relatively quiescent solar activity. However, the 12\% discrepancy at the brighter end of R4, R5, and R6 is not well understood. Considering the different instrument design and space environments, the origin of the (small) discrepancy between the coefficient and a 1:1 relation might be convoluted as well. This systematic difference is small, however, and we speculate that it is accounted for by including calibration differences, which were not tested in this work.

In Fig.~\ref{fig:uncertainty} we show the rate comparison in the R4 band in detail (top panels). They show no obvious offset at the zeropoint. The resemblance of the data from the two observations is seen down to very low rates ($\rm \simeq 0.05\, cts\, s^{-1}\, deg^{-2}$), corroborating the quality of the systematic background calibration at this extent.
In the bottom panels of Fig.\ref{fig:uncertainty}, we show the scatter of the \rosat{} data. By taking a slice of pixels at about $\rm 0.53\, cts\, s^{-1}\, deg^{-2}$ of the converted \erosita{} rates (cyan area in the top right panel), we draw the distribution of their \rosat{} rates (bottom left panel) and compute the full width at half maximum (FWHM) of the distribution. In general, in two observations detecting the same flux with high statistics and no systematics (i.e., Poisson uncertainty only), the rate distribution should be a Gaussian with FWHM $=2.3548~\sigma$, where $\sigma$ is the standard deviation of the Gaussian. We thus wish to compare the width of the distribution, converted into $\sigma$, to the intrinsic \rosat{} uncertainty. In the bottom right panel of Fig.\ref{fig:uncertainty}, we show the measured $\sigma$ at fixed \erosita{} equivalent rates for all rates. We note that the $\sigma$ (solid blue line) grows together with the \rosat{} uncertainty (dashed black line). We find that $\sigma$ is about 1.25 times of the \rosat{} uncertainty for all bands (see the dotted black line). The \erosita{} uncertainty was ignored in this computation as negligible compared to the \rosat{} uncertainty. According to this comparison, the \erosita{} data agree with the \rosat{} within 1.25 times the \rosat{} uncertainty. They are thus consistent.

\section{Large-scale sources of diffuse emission} \label{sec:features}

\subsection{The case of the \erosita\ bubbles}\label{sec:erobubbles}

\begin{figure*}[t!]
    \includegraphics[width=0.95\textwidth,trim=-20 0 20 0, clip]{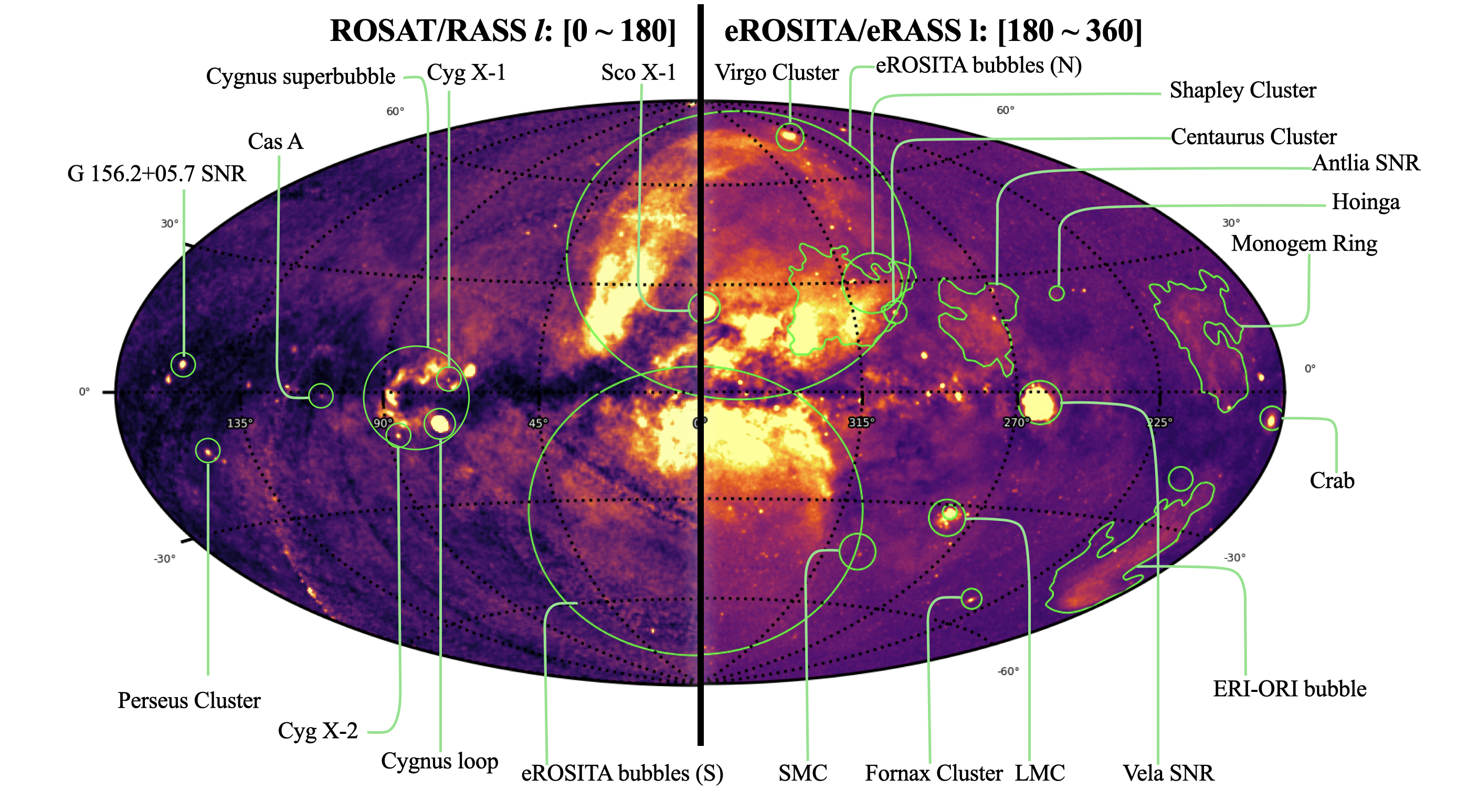}
    \caption{Finding chart of the all-sky diffuse emission.
    The western sky $l$=[180, 360] is shown with the \erosita{}/eRASS1 data in the 0.4--0.6 keV band. 
    The eastern sky $l$=[0, 180] is shown with the \rosat{}/RASS R4 (0.44--1.01 keV) band map.
    The green marks show the location of large projected structures, including known supernova remnants, the Large and Small Magellanic Clouds, and some extended clusters and galaxy superclusters. The color bar has been arbitrarily chosen to match the dynamic ranges of the maps.
    }
    \label{fig:find_chart}
\end{figure*}

Because of the extraordinary concordance between the \rosat\ and \erosita\ maps, it is reasonable to question why the \erosita\ bubbles were not immediately recognized on the basis of the \rosat\ maps \citep{Snowden1997ApJ}. We note that the main features that allowed \citet{Predehl2020Natur} to recognize the \erosita\ bubbles are their hourglass shape, which comes from the fourfold symmetry around the Galactic center, which comes about through the detection of faint diffuse emission at high Galactic latitudes. 

Based on the \rosat\ maps, the existence of a bright X-ray spur northeast of the Galactic center was clearly recognized, which is now known as the North Polar Spur \citep{Iwan1980ApJ, Sofue1979AA}. By virtue of the \erosita\ maps, it is evident that the continuation of the North Polar Spur forms a loop, with a fainter edge on the northwestern side of the Galactic center \citep{Predehl2020Natur}. Considerable complexity along the northwestern edge of the bubbles was already recognized based on the \rosat\ maps in particular. Through the attempt of decomposition for the diffuse \rosat\ emission in \citet{Freyberg1994PhDT}, some emission features from local diffuse features were possibly separated. Fig.\ref{fig:Michael2} shows a feature at $\rm l,b \sim (315, 20)$ that was initially attributed to a local SNR. Even in the higher-quality \erosita\ maps, this foreground emission complicates the identification of the northwestern edge of the \erosita\ bubbles, which is well defined, but rather faint at high Galactic latitudes ($b\ge30^\circ$), while its edge becomes unclear at lower latitudes. Therefore, it is not surprising that the northwest edge of the bubble was not recognized in the \rosat{} data.

The southwest edge of the \erosita\ bubbles is instead much clearer. They can clearly be delineated from few degrees below the Galactic disk all the way to $|b|\sim50^{\degree}$ \citep{Predehl2020Natur}. However, the very clear edge is severely confused in the \rosat\ map. This location on the sky corresponds to the passage of \rosat\ through the South Atlantic Anomaly, which is characterized by an elevated background and induces a significantly shallower exposure in the \rosat\ maps at that location (see Fig.\ref{fig:Michael2}). Therefore, it is not surprising that the faint southeast edge of the \erosita\ bubbles has not been recognized in the lower-sensitivity \rosat\ maps. 

Considering that the southeast edge of the \erosita\ bubbles was not detected because of its faintness, its southwest edge was not noticed because of the higher background and shorter exposure related with the passage of \rosat\ through the South Atlantic Anomaly, its northwest edge was confused by large-scale features (which was attributed to a local supernova remnant) and the North Polar Spur was thought to be a uniquely a local feature, it is not surprising that the footprint of the \erosita\ bubbles could not be recognized on the basis of the \rosat\ maps. However, studies \citep{Kataoka2018Galax} have already anticipated the existence of the \erosita{} bubbles with high certainty from the gamma-ray observation.

\subsection{Other sources}

Direct visual examination of the eRASS1 maps revealed a variety of structures of diffuse emission. In this section, we enumerate the prominent diffuse emission features and structures in the maps and briefly characterize them in order to provide an idea of the diffuse emission sky. For this purpose, Fig.\ref{fig:find_chart} shows the locations and extent of large projected coherent patches of diffuse emission in the soft X-ray band. In the eastern Galactic hemisphere ($l$=[0, 180]), we show the \rosat{} R4 map of 0.44--1.01 keV, and in the western Galactic hemisphere ($l$=[180, 360]), we show the \erosita{} data in the 0.4--0.6 keV band.

Within the Galactic plane, only local emission ($\leq 200$ pc) is not damped by the high absorption. At intermediate Galactic latitudes, large-scale diffuse emission generated by the hot gas in the halo of the Milky Way and supernova remnants becomes noticeable. Large-scale features such as the Monogem Ring at $l,b\sim$ (200, 15) \citep[][Knies et al. in prep.]{Knies2018MNRAS}, the Orion-Eridanus superbubble at $l,b\sim$ (210, -40), an expanding superbubble filled with X-ray emitting plasma rising from the Galactic plane \citep{Joubaud2019AA,Burrows1993ApJ,Heiles1999ASPC}, the Antlia supernova remnant at $l,b\sim$ (280, 23) (Knies et al. in prep.), the recently discovered Hoinga at $l,b\sim$ (249.5, 24.5) \citep{Becker2021AA}, and the nearby SNR $l,b\sim$ (320, 20) are clearly distinguished.

The comparison of \erosita{} maps with the corresponding \rosat{} maps additionally reveals brightness and/or flux changes for variable X-ray sources, such as novae and X-ray binaries. For example, the X-ray transient nova Musca 1991 (G295-7) was evident in the \rosat{} maps through all bands, but it is not detectable (or is significantly fainter) in the \erosita{} maps (see also Fig.\ref{fig:ero_r4}--Fig.\ref{fig:ero_r7}).

\section{Summary}\label{sec:summary}

We presented the diffuse X-ray emission maps of the eRASS1 data in different energy bands from 0.2 to 8.0 keV. We demonstrated the capability of \erosita{} to detect faint diffuse emission sources, including the LHB, the Galactic halo, various SNRs, the ISM, and the CXB through soft to hard energies. We reported the observed rates, surface brightness, and total flux in the western Galactic hemisphere. The energy resolution of \erosita{} allows a detailed investigation of the Galactic diffuse plasma, with a narrower energy selection than for the \rosat{} data.

The data from \erosita{}/eRASS1 were quantitatively compared with the \rosat{} maps in the \rosat{} standard bands, after taking into account the instrumental background and response efficiency. The visual inspection of eRASS1 and RASS maps already shows strong similarities in the morphology of the diffuse emission in the whole hemisphere. Through a quantitative comparison, the fit of the densest data sample shows that \erosita{} observes count rates consistent with \rosat{}, with a weak systematic tendency to lower rates. The correlation coefficient is $\sim90\%$ in R4, R5, and R6, and it is $\sim 99\%$ in R7, while the scatter of the data is about 1.25 times of the measurement uncertainty in all bands. We attribute the small residual deviation from the 1:1 relation as potentially caused by the different calibration procedures and by the subtraction of emission induced by solar activity. We sketched the all-sky atlas of the prominent diffuse emission structures present in the literature and detected by \erosita{} and provided an overview of the known diffuse sources in the soft X-ray sky.

The following \erosita{} all-sky surveys will provide more details for sources of diffuse emission, allowing us to improve the characterization of the various plasma components and environment of the Milky Way, nearby galaxies, and the extragalactic sky. 
The broadband data products presented in this work will soon be made available in a dedicated \erosita{} repository during the eRASS1 data release.

\begin{acknowledgements}
This work is based on data from eROSITA, the soft X-ray instrument aboard SRG, a joint Russian-German science mission supported by the Russian Space Agency (Roskosmos), in the interests of the Russian Academy of Sciences represented by its Space Research Institute (IKI), and the Deutsches Zentrum für Luft- und Raumfahrt (DLR). The SRG spacecraft was built by Lavochkin Association (NPOL) and its subcontractors, and is operated by NPOL with support from the Max Planck Institute for Extraterrestrial Physics (MPE). The development and construction of the eROSITA X-ray instrument was led by MPE, with contributions from the Dr. Karl Remeis Observatory Bamberg \& ECAP (FAU Erlangen-Nuernberg), the University of Hamburg Observatory, the Leibniz Institute for Astrophysics Potsdam (AIP), and the Institute for Astronomy and Astrophysics of the University of Tübingen, with the support of DLR and the Max Planck Society. The Argelander Institute for Astronomy of the University of Bonn and the Ludwig Maximilians Universität Munich also participated in the science preparation for eROSITA. The eROSITA data shown here were processed using the eSASS software system developed by the German eROSITA consortium. This research made use of Astropy,\footnote{http://www.astropy.org} a community-developed core Python package for Astronomy \citep{astropy:2013, astropy:2018}. Some of the results in this paper have been derived using the healpy and HEALPix packages. We acknowledge financial support from the European Research Council (ERC) under the European Union’s Horizon 2020 research and innovation program Hot- Milk (grant agreement No 865637)
\end{acknowledgements}

\bibliography{aa}

\begin{thebibliography}{49}
\expandafter\ifx\csname natexlab\endcsname\relax\def\natexlab#1{#1}\fi

\bibitem[{{Astropy Collaboration} {et~al.}(2018){Astropy Collaboration}, {Price-Whelan}, {Sip{\H{o}}cz}, {G{\"u}nther}, {Lim}, {Crawford}, {Conseil}, {Shupe}, {Craig}, {Dencheva}, {Ginsburg}, {VanderPlas}, {Bradley}, {P{\'e}rez-Su{\'a}rez}, {de Val-Borro}, {Aldcroft}, {Cruz}, {Robitaille}, {Tollerud}, {Ardelean}, {Babej}, {Bach}, {Bachetti}, {Bakanov}, {Bamford}, {Barentsen}, {Barmby}, {Baumbach}, {Berry}, {Biscani}, {Boquien}, {Bostroem}, {Bouma}, {Brammer}, {Bray}, {Breytenbach}, {Buddelmeijer}, {Burke}, {Calderone}, {Cano Rodr{\'\i}guez}, {Cara}, {Cardoso}, {Cheedella}, {Copin}, {Corrales}, {Crichton}, {D'Avella}, {Deil}, {Depagne}, {Dietrich}, {Donath}, {Droettboom}, {Earl}, {Erben}, {Fabbro}, {Ferreira}, {Finethy}, {Fox}, {Garrison}, {Gibbons}, {Goldstein}, {Gommers}, {Greco}, {Greenfield}, {Groener}, {Grollier}, {Hagen}, {Hirst}, {Homeier}, {Horton}, {Hosseinzadeh}, {Hu}, {Hunkeler}, {Ivezi{\'c}}, {Jain}, {Jenness}, {Kanarek}, {Kendrew}, {Kern}, {Kerzendorf}, {Khvalko}, {King}, {Kirkby}, {Kulkarni},
  {Kumar}, {Lee}, {Lenz}, {Littlefair}, {Ma}, {Macleod}, {Mastropietro}, {McCully}, {Montagnac}, {Morris}, {Mueller}, {Mumford}, {Muna}, {Murphy}, {Nelson}, {Nguyen}, {Ninan}, {N{\"o}the}, {Ogaz}, {Oh}, {Parejko}, {Parley}, {Pascual}, {Patil}, {Patil}, {Plunkett}, {Prochaska}, {Rastogi}, {Reddy Janga}, {Sabater}, {Sakurikar}, {Seifert}, {Sherbert}, {Sherwood-Taylor}, {Shih}, {Sick}, {Silbiger}, {Singanamalla}, {Singer}, {Sladen}, {Sooley}, {Sornarajah}, {Streicher}, {Teuben}, {Thomas}, {Tremblay}, {Turner}, {Terr{\'o}n}, {van Kerkwijk}, {de la Vega}, {Watkins}, {Weaver}, {Whitmore}, {Woillez}, {Zabalza}, \& {Astropy Contributors}}]{astropy:2018}
{Astropy Collaboration}, {Price-Whelan}, A.~M., {Sip{\H{o}}cz}, B.~M., {et~al.} 2018, \href{http://dx.doi.org/10.3847/1538-3881/aabc4f}{\color{magenta}\aj}, \href{https://ui.adsabs.harvard.edu/abs/2018AJ....156..123A}{156, 123}

\bibitem[{{Astropy Collaboration} {et~al.}(2013){Astropy Collaboration}, {Robitaille}, {Tollerud}, {Greenfield}, {Droettboom}, {Bray}, {Aldcroft}, {Davis}, {Ginsburg}, {Price-Whelan}, {Kerzendorf}, {Conley}, {Crighton}, {Barbary}, {Muna}, {Ferguson}, {Grollier}, {Parikh}, {Nair}, {Unther}, {Deil}, {Woillez}, {Conseil}, {Kramer}, {Turner}, {Singer}, {Fox}, {Weaver}, {Zabalza}, {Edwards}, {Azalee Bostroem}, {Burke}, {Casey}, {Crawford}, {Dencheva}, {Ely}, {Jenness}, {Labrie}, {Lim}, {Pierfederici}, {Pontzen}, {Ptak}, {Refsdal}, {Servillat}, \& {Streicher}}]{astropy:2013}
{Astropy Collaboration}, {Robitaille}, T.~P., {Tollerud}, E.~J., {et~al.} 2013, \href{http://dx.doi.org/10.1051/0004-6361/201322068}{\color{magenta}\aap}, \href{https://ui.adsabs.harvard.edu/abs/2013A&A...558A..33A}{558, A33}

\bibitem[{{Balucinska-Church} \& {McCammon}(1992)}]{Balucinska1992ApJ}
{Balucinska-Church}, M. \& {McCammon}, D. 1992, \href{http://dx.doi.org/10.1086/172032}{\color{magenta}\apj}, \href{https://ui.adsabs.harvard.edu/abs/1992ApJ...400..699B}{400, 699}

\bibitem[{{Becker} {et~al.}(2021){Becker}, {Hurley-Walker}, {Weinberger}, {Nicastro}, {Mayer}, {Merloni}, \& {Sanders}}]{Becker2021AA}
{Becker}, W., {Hurley-Walker}, N., {Weinberger}, C., {et~al.} 2021, \href{http://dx.doi.org/10.1051/0004-6361/202040156}{\color{magenta}\aap}, \href{https://ui.adsabs.harvard.edu/abs/2021A&A...648A..30B}{648, A30}

\bibitem[{{Brandt} \& {Yang}(2021)}]{Brant2021arXiv211101156B}
{Brandt}, W.~N. \& {Yang}, G. 2021, \href{https://ui.adsabs.harvard.edu/abs/2021arXiv211101156B}{arXiv e-prints, arXiv:2111.01156}

\bibitem[{{Brunner} {et~al.}(2022){Brunner}, {Liu}, {Lamer}, {Georgakakis}, {Merloni}, {Brusa}, {Bulbul}, {Dennerl}, {Friedrich}, {Liu}, {Maitra}, {Nandra}, {Ramos-Ceja}, {Sanders}, {Stewart}, {Boller}, {Buchner}, {Clerc}, {Comparat}, {Dwelly}, {Eckert}, {Finoguenov}, {Freyberg}, {Ghirardini}, {Gueguen}, {Haberl}, {Kreykenbohm}, {Krumpe}, {Osterhage}, {Pacaud}, {Predehl}, {Reiprich}, {Robrade}, {Salvato}, {Santangelo}, {Schrabback}, {Schwope}, \& {Wilms}}]{Brunner2022AA}
{Brunner}, H., {Liu}, T., {Lamer}, G., {et~al.} 2022, \href{http://dx.doi.org/10.1051/0004-6361/202141266}{\color{magenta}\aap}, \href{https://ui.adsabs.harvard.edu/abs/2022A&A...661A...1B}{661, A1}

\bibitem[{{Burrows} {et~al.}(1993){Burrows}, {Singh}, {Nousek}, {Garmire}, \& {Good}}]{Burrows1993ApJ}
{Burrows}, D.~N., {Singh}, K.~P., {Nousek}, J.~A., {Garmire}, G.~P., \& {Good}, J. 1993, \href{http://dx.doi.org/10.1086/172423}{\color{magenta}\apj}, \href{https://ui.adsabs.harvard.edu/abs/1993ApJ...406...97B}{406, 97}

\bibitem[{{Freyberg}(1994)}]{Freyberg1994PhDT}
{Freyberg}, M. 1994, \href{https://ui.adsabs.harvard.edu/abs/1994PhDT.......103F}{{Untersuchungen der kosmischen und nichtkosmischen Komponenten der R{\"o}ntgenhintergrundstrahlung mit ROSAT}}, PhD thesis, Ludwig-Maximilians University of Munich, Germany

\bibitem[{{Freyberg} {et~al.}(2020){Freyberg}, {Perinati}, {Pacaud}, {Eraerds}, {Churazov}, {Dennerl}, {Predehl}, {Merloni}, {Meidinger}, {Bulbul}, {Friedrich}, {Gilfanov}, {Tenzer}, {Pommranz}, {Eckert}, {Schmitt}, {Brusa}, \& {Santangelo}}]{Freyberg2020SPIE}
{Freyberg}, M., {Perinati}, E., {Pacaud}, F., {et~al.} 2020, in Society of Photo-Optical Instrumentation Engineers (SPIE) Conference Series, Vol. 11444, Society of Photo-Optical Instrumentation Engineers (SPIE) Conference Series, \href{https://ui.adsabs.harvard.edu/abs/2020SPIE11444E..1OF}{114441O}

\bibitem[{{Gilli} {et~al.}(2007){Gilli}, {Comastri}, \& {Hasinger}}]{Gilli2007AA}
{Gilli}, R., {Comastri}, A., \& {Hasinger}, G. 2007, \href{http://dx.doi.org/10.1051/0004-6361:20066334}{\color{magenta}\aap}, \href{https://ui.adsabs.harvard.edu/abs/2007A&A...463...79G}{463, 79}

\bibitem[{{G{\'o}rski} {et~al.}(2005){G{\'o}rski}, {Hivon}, {Banday}, {Wandelt}, {Hansen}, {Reinecke}, \& {Bartelmann}}]{Healpy2005ApJ}
{G{\'o}rski}, K.~M., {Hivon}, E., {Banday}, A.~J., {et~al.} 2005, \href{http://dx.doi.org/10.1086/427976}{\color{magenta}\apj}, \href{http://adsabs.harvard.edu/abs/2005ApJ...622..759G}{622, 759}

\bibitem[{{Heiles} {et~al.}(1999){Heiles}, {Haffner}, \& {Reynolds}}]{Heiles1999ASPC}
{Heiles}, C., {Haffner}, L.~M., \& {Reynolds}, R.~J. 1999, in Astronomical Society of the Pacific Conference Series, Vol. 168, New Perspectives on the Interstellar Medium, ed. A.~R. {Taylor}, T.~L. {Landecker}, \& G.~{Joncas}, \href{https://ui.adsabs.harvard.edu/abs/1999ASPC..168..211H}{211}

\bibitem[{{HI4PI Collaboration} {et~al.}(2016){HI4PI Collaboration}, {Ben Bekhti}, {Fl{\"o}er}, {Keller}, {Kerp}, {Lenz}, {Winkel}, {Bailin}, {Calabretta}, {Dedes}, {Ford}, {Gibson}, {Haud}, {Janowiecki}, {Kalberla}, {Lockman}, {McClure-Griffiths}, {Murphy}, {Nakanishi}, {Pisano}, \& {Staveley-Smith}}]{HI4PI2016AA}
{HI4PI Collaboration}, {Ben Bekhti}, N., {Fl{\"o}er}, L., {et~al.} 2016, \href{http://dx.doi.org/10.1051/0004-6361/201629178}{\color{magenta}\aap}, \href{https://ui.adsabs.harvard.edu/abs/2016A&A...594A.116H}{594, A116}

\bibitem[{{Iwan}(1980)}]{Iwan1980ApJ}
{Iwan}, D. 1980, \href{http://dx.doi.org/10.1086/158113}{\color{magenta}\apj}, \href{https://ui.adsabs.harvard.edu/abs/1980ApJ...239..316I}{239, 316}

\bibitem[{{Jin} {et~al.}(2017){Jin}, {Ponti}, {Haberl}, \& {Smith}}]{Jin2017MNRAS}
{Jin}, C., {Ponti}, G., {Haberl}, F., \& {Smith}, R. 2017, \href{http://dx.doi.org/10.1093/mnras/stx653}{\color{magenta}\mnras}, \href{https://ui.adsabs.harvard.edu/abs/2017MNRAS.468.2532J}{468, 2532}

\bibitem[{{Joubaud} {et~al.}(2019){Joubaud}, {Grenier}, {Ballet}, \& {Soler}}]{Joubaud2019AA}
{Joubaud}, T., {Grenier}, I.~A., {Ballet}, J., \& {Soler}, J.~D. 2019, \href{http://dx.doi.org/10.1051/0004-6361/201936239}{\color{magenta}\aap}, \href{https://ui.adsabs.harvard.edu/abs/2019A&A...631A..52J}{631, A52}

\bibitem[{{Kataoka} {et~al.}(2018){Kataoka}, {Sofue}, {Inoue}, {Akita}, {Nakashima}, \& {Totani}}]{Kataoka2018Galax}
{Kataoka}, J., {Sofue}, Y., {Inoue}, Y., {et~al.} 2018, \href{http://dx.doi.org/10.3390/galaxies6010027}{\color{magenta}Galaxies}, \href{https://ui.adsabs.harvard.edu/abs/2018Galax...6...27K}{6, 27}

\bibitem[{{Kerp} {et~al.}(1999){Kerp}, {Burton}, {Egger}, {Freyberg}, {Hartmann}, {Kalberla}, {Mebold}, \& {Pietz}}]{Kerp1999AA}
{Kerp}, J., {Burton}, W.~B., {Egger}, R., {et~al.} 1999, \href{http://dx.doi.org/10.48550/arXiv.astro-ph/9810307}{\color{magenta}\aap}, \href{https://ui.adsabs.harvard.edu/abs/1999A&A...342..213K}{342, 213}

\bibitem[{{Knies} {et~al.}(2018){Knies}, {Sasaki}, \& {Plucinsky}}]{Knies2018MNRAS}
{Knies}, J.~R., {Sasaki}, M., \& {Plucinsky}, P.~P. 2018, \href{http://dx.doi.org/10.1093/mnras/sty915}{\color{magenta}\mnras}, \href{https://ui.adsabs.harvard.edu/abs/2018MNRAS.477.4414K}{477, 4414}

\bibitem[{{Koutroumpa} {et~al.}(2006){Koutroumpa}, {Lallement}, {Kharchenko}, {Dalgarno}, {Pepino}, {Izmodenov}, \& {Qu{\'e}merais}}]{Koutroumpa2006AA}
{Koutroumpa}, D., {Lallement}, R., {Kharchenko}, V., {et~al.} 2006, \href{http://dx.doi.org/10.1051/0004-6361:20065250}{\color{magenta}\aap}, \href{https://ui.adsabs.harvard.edu/abs/2006A&A...460..289K}{460, 289}

\bibitem[{{Lallement} {et~al.}(2018){Lallement}, {Capitanio}, {Ruiz-Dern}, {Danielski}, {Babusiaux}, {Vergely}, {Elyajouri}, {Arenou}, \& {Leclerc}}]{Lallement2018AA}
{Lallement}, R., {Capitanio}, L., {Ruiz-Dern}, L., {et~al.} 2018, \href{http://dx.doi.org/10.1051/0004-6361/201832832}{\color{magenta}\aap}, \href{https://ui.adsabs.harvard.edu/abs/2018A&A...616A.132L}{616, A132}

\bibitem[{{Lallement} {et~al.}(2016){Lallement}, {Snowden}, {Kuntz}, {Dame}, {Koutroumpa}, {Grenier}, \& {Casandjian}}]{Lallement2016AA}
{Lallement}, R., {Snowden}, S., {Kuntz}, K.~D., {et~al.} 2016, \href{http://dx.doi.org/10.1051/0004-6361/201629453}{\color{magenta}\aap}, \href{https://ui.adsabs.harvard.edu/abs/2016A&A...595A.131L}{595, A131}

\bibitem[{{Lallement} {et~al.}(2015){Lallement}, {Vergely}, {Puspitarini}, {Snowden}, {Galeazzi}, \& {Koutroumpa}}]{Lallement2015MmSAI}
{Lallement}, R., {Vergely}, J.~L., {Puspitarini}, L., {et~al.} 2015, \memsai, \href{https://ui.adsabs.harvard.edu/abs/2015MmSAI..86..626L}{86, 626}

\bibitem[{{Lamer} {et~al.}(2021){Lamer}, {Schwope}, {Predehl}, {Traulsen}, {Wilms}, \& {Freyberg}}]{Lamer2021AA}
{Lamer}, G., {Schwope}, A.~D., {Predehl}, P., {et~al.} 2021, \href{http://dx.doi.org/10.1051/0004-6361/202039757}{\color{magenta}\aap}, \href{https://ui.adsabs.harvard.edu/abs/2021A&A...647A...7L}{647, A7}

\bibitem[{{Liu} {et~al.}(2017){Liu}, {Chiao}, {Collier}, {Cravens}, {Galeazzi}, {Koutroumpa}, {Kuntz}, {Lallement}, {Lepri}, {McCammon}, {Morgan}, {Porter}, {Snowden}, {Thomas}, {Uprety}, {Ursino}, \& {Walsh}}]{Liu2017ApJ}
{Liu}, W., {Chiao}, M., {Collier}, M.~R., {et~al.} 2017, \href{http://dx.doi.org/10.3847/1538-4357/834/1/33}{\color{magenta}\apj}, \href{https://ui.adsabs.harvard.edu/abs/2017ApJ...834...33L}{834, 33}

\bibitem[{{Locatelli} {et~al.}(2022){Locatelli}, {Ponti}, \& {Bianchi}}]{Locatelli2022AA}
{Locatelli}, N., {Ponti}, G., \& {Bianchi}, S. 2022, \href{http://dx.doi.org/10.1051/0004-6361/202142655}{\color{magenta}\aap}, \href{https://ui.adsabs.harvard.edu/abs/2022A&A...659A.118L}{659, A118}

\bibitem[{{Locatelli} {et~al.}(2023){Locatelli}, {Ponti}, {Zheng}, {Merloni}, {Becker}, {Comparat}, {Dennerl}, {Freyberg}, {Sasaki}, \& {Yeung}}]{2023arXiv231010715L}
{Locatelli}, N., {Ponti}, G., {Zheng}, X., {et~al.} 2023, \href{https://ui.adsabs.harvard.edu/abs/2023arXiv231010715L}{\href{http://dx.doi.org/10.48550/arXiv.2310.10715}{\color{magenta}arXiv e-prints}, arXiv:2310.10715}

\bibitem[{{McCammon} {et~al.}(2002){McCammon}, {Almy}, {Apodaca}, {Bergmann Tiest}, {Cui}, {Deiker}, {Galeazzi}, {Juda}, {Lesser}, {Mihara}, {Morgenthaler}, {Sanders}, {Zhang}, {Figueroa-Feliciano}, {Kelley}, {Moseley}, {Mushotzky}, {Porter}, {Stahle}, \& {Szymkowiak}}]{McCammon2002ApJ}
{McCammon}, D., {Almy}, R., {Apodaca}, E., {et~al.} 2002, \href{http://dx.doi.org/10.1086/341727}{\color{magenta}\apj}, \href{https://ui.adsabs.harvard.edu/abs/2002ApJ...576..188M}{576, 188}

\bibitem[{{Merloni} {et~al.}(2023){Merloni}, {Lamer}, {Liu}, \& {Ramos-Ceja}}]{Merloni23}
{Merloni}, A., {Lamer}, G., {Liu}, T., \& {Ramos-Ceja}, M.~E. 2023, \aap, submitted, submitted

\bibitem[{{Merloni} {et~al.}(2012){Merloni}, {Predehl}, {Becker}, {B{\"o}hringer}, {Boller}, {Brunner}, {Brusa}, {Dennerl}, {Freyberg}, {Friedrich}, {Georgakakis}, {Haberl}, {Hasinger}, {Meidinger}, {Mohr}, {Nandra}, {Rau}, {Reiprich}, {Robrade}, {Salvato}, {Santangelo}, {Sasaki}, {Schwope}, {Wilms}, \& {the German eROSITA Consortium}}]{erositabook}
{Merloni}, A., {Predehl}, P., {Becker}, W., {et~al.} 2012, \href{https://ui.adsabs.harvard.edu/abs/2012arXiv1209.3114M}{arXiv e-prints, arXiv:1209.3114}

\bibitem[{{Pietz} {et~al.}(1998){Pietz}, {Kerp}, {Kalberla}, {Burton}, {Hartmann}, \& {Mebold}}]{Pietz1998AA}
{Pietz}, J., {Kerp}, J., {Kalberla}, P.~M.~W., {et~al.} 1998, \aap, \href{https://ui.adsabs.harvard.edu/abs/1998A&A...332...55P}{332, 55}

\bibitem[{{Ponti} {et~al.}(2023){Ponti}, {Zheng}, {Locatelli}, {Bianchi}, {Zhang}, {Anastasopoulou}, {Comparat}, {Dennerl}, {Freyberg}, {Haberl}, {Merloni}, {Reiprich}, {Salvato}, {Sanders}, {Sasaki}, {Strong}, \& {Yeung}}]{Ponti2023AA1}
{Ponti}, G., {Zheng}, X., {Locatelli}, N., {et~al.} 2023, \href{http://dx.doi.org/10.1051/0004-6361/202243992}{\color{magenta}\aap}, \href{https://ui.adsabs.harvard.edu/abs/2023A&A...674A.195P}{674, A195}

\bibitem[{{Predehl} {et~al.}(2021){Predehl}, {Andritschke}, {Arefiev}, {Babyshkin}, {Batanov}, {Becker}, {B{\"o}hringer}, {Bogomolov}, {Boller}, {Borm}, {Bornemann}, {Br{\"a}uninger}, {Br{\"u}ggen}, {Brunner}, {Brusa}, {Bulbul}, {Buntov}, {Burwitz}, {Burkert}, {Clerc}, {Churazov}, {Coutinho}, {Dauser}, {Dennerl}, {Doroshenko}, {Eder}, {Emberger}, {Eraerds}, {Finoguenov}, {Freyberg}, {Friedrich}, {Friedrich}, {F{\"u}rmetz}, {Georgakakis}, {Gilfanov}, {Granato}, {Grossberger}, {Gueguen}, {Gureev}, {Haberl}, {H{\"a}lker}, {Hartner}, {Hasinger}, {Huber}, {Ji}, {Kienlin}, {Kink}, {Korotkov}, {Kreykenbohm}, {Lamer}, {Lomakin}, {Lapshov}, {Liu}, {Maitra}, {Meidinger}, {Menz}, {Merloni}, {Mernik}, {Mican}, {Mohr}, {M{\"u}ller}, {Nandra}, {Nazarov}, {Pacaud}, {Pavlinsky}, {Perinati}, {Pfeffermann}, {Pietschner}, {Ramos-Ceja}, {Rau}, {Reiffers}, {Reiprich}, {Robrade}, {Salvato}, {Sanders}, {Santangelo}, {Sasaki}, {Scheuerle}, {Schmid}, {Schmitt}, {Schwope}, {Shirshakov}, {Steinmetz}, {Stewart}, {Str{\"u}der},
  {Sunyaev}, {Tenzer}, {Tiedemann}, {Tr{\"u}mper}, {Voron}, {Weber}, {Wilms}, \& {Yaroshenko}}]{Predehl2021AA}
{Predehl}, P., {Andritschke}, R., {Arefiev}, V., {et~al.} 2021, \href{http://dx.doi.org/10.1051/0004-6361/202039313}{\color{magenta}\aap}, \href{https://ui.adsabs.harvard.edu/abs/2021A&A...647A...1P}{647, A1}

\bibitem[{{Predehl} {et~al.}(2020{\natexlab{a}}){Predehl}, {Andritschke}, {Arefiev}, {Babyshkin}, {Batanov}, {Becker}, {B{\"o}hringer}, {Bogomolov}, {Boller}, {Borm}, {Bornemann}, {Br{\"a}uninger}, {Br{\"u}ggen}, {Brunner}, {Brusa}, {Bulbul}, {Buntov}, {Burwitz}, {Burkert}, {Clerc}, {Churazov}, {Coutinho}, {Dauser}, {Dennerl}, {Doroshenko}, {Eder}, {Emberger}, {Eraerds}, {Finoguenov}, {Freyberg}, {Friedrich}, {Friedrich}, {F{\"u}rmetz}, {Georgakakis}, {Gilfanov}, {Granato}, {Grossberger}, {Gueguen}, {Gureev}, {Haberl}, {H{\"a}lker}, {Hartner}, {Hasinger}, {Huber}, {Ji}, {Kienlin}, {Kink}, {Korotkov}, {Kreykenbohm}, {Lamer}, {Lomakin}, {Lapshov}, {Liu}, {Maitra}, {Meidinger}, {Menz}, {Merloni}, {Mernik}, {Mican}, {Mohr}, {M{\"u}ller}, {Nandra}, {Nazarov}, {Pacaud}, {Pavlinsky}, {Perinati}, {Pfeffermann}, {Pietschner}, {Ramos-Ceja}, {Rau}, {Reiffers}, {Reiprich}, {Robrade}, {Salvato}, {Sanders}, {Santangelo}, {Sasaki}, {Scheuerle}, {Schmid}, {Schmitt}, {Schwope}, {Shirshakov}, {Steinmetz}, {Stewart},
  {Str{\"u}der}, {Sunyaev}, {Tenzer}, {Tiedemann}, {Tr{\"u}mper}, {Voron}, {Weber}, {Wilms}, \& {Yaroshenko}}]{2020Predehl_instr}
{Predehl}, P., {Andritschke}, R., {Arefiev}, V., {et~al.} 2020{\natexlab{a}}, \href{https://ui.adsabs.harvard.edu/abs/2020arXiv201003477P}{arXiv e-prints, arXiv:2010.03477}

\bibitem[{{Predehl} \& {Schmitt}(1995)}]{Predehl1995AA}
{Predehl}, P. \& {Schmitt}, J.~H.~M.~M. 1995, \aap, \href{https://ui.adsabs.harvard.edu/abs/1995A&A...293..889P}{500, 459}

\bibitem[{{Predehl} {et~al.}(2020{\natexlab{b}}){Predehl}, {Sunyaev}, {Becker}, {Brunner}, {Burenin}, {Bykov}, {Cherepashchuk}, {Chugai}, {Churazov}, {Doroshenko}, {Eismont}, {Freyberg}, {Gilfanov}, {Haberl}, {Khabibullin}, {Krivonos}, {Maitra}, {Medvedev}, {Merloni}, {Nandra}, {Nazarov}, {Pavlinsky}, {Ponti}, {Sanders}, {Sasaki}, {Sazonov}, {Strong}, \& {Wilms}}]{Predehl2020Natur}
{Predehl}, P., {Sunyaev}, R.~A., {Becker}, W., {et~al.} 2020{\natexlab{b}}, \href{http://dx.doi.org/10.1038/s41586-020-2979-0}{\color{magenta}\nat}, \href{https://ui.adsabs.harvard.edu/abs/2020Natur.588..227P}{588, 227}

\bibitem[{{Reid} \& {Brunthaler}(2004)}]{Reid2004ApJ}
{Reid}, M.~J. \& {Brunthaler}, A. 2004, \href{http://dx.doi.org/10.1086/424960}{\color{magenta}\apj}, \href{https://ui.adsabs.harvard.edu/abs/2004ApJ...616..872R}{616, 872}

\bibitem[{{Snowden} {et~al.}(1998){Snowden}, {Egger}, {Finkbeiner}, {Freyberg}, \& {Plucinsky}}]{Snowden1998ApJ}
{Snowden}, S.~L., {Egger}, R., {Finkbeiner}, D.~P., {Freyberg}, M.~J., \& {Plucinsky}, P.~P. 1998, \href{http://dx.doi.org/10.1086/305135}{\color{magenta}\apj}, \href{https://ui.adsabs.harvard.edu/abs/1998ApJ...493..715S}{493, 715}

\bibitem[{{Snowden} {et~al.}(1997){Snowden}, {Egger}, {Freyberg}, {McCammon}, {Plucinsky}, {Sanders}, {Schmitt}, {Tr{\"u}mper}, \& {Voges}}]{Snowden1997ApJ}
{Snowden}, S.~L., {Egger}, R., {Freyberg}, M.~J., {et~al.} 1997, \href{http://dx.doi.org/10.1086/304399}{\color{magenta}\apj}, \href{https://ui.adsabs.harvard.edu/abs/1997ApJ...485..125S}{485, 125}

\bibitem[{{Snowden} \& {Freyberg}(1993)}]{Snowden1993ApJ}
{Snowden}, S.~L. \& {Freyberg}, M.~J. 1993, \href{http://dx.doi.org/10.1086/172289}{\color{magenta}\apj}, \href{https://ui.adsabs.harvard.edu/abs/1993ApJ...404..403S}{404, 403}

\bibitem[{{Snowden} {et~al.}(2000){Snowden}, {Freyberg}, {Kuntz}, \& {Sanders}}]{Snowden2000ApJS}
{Snowden}, S.~L., {Freyberg}, M.~J., {Kuntz}, K.~D., \& {Sanders}, W.~T. 2000, \href{http://dx.doi.org/10.1086/313378}{\color{magenta}\apjs}, \href{https://ui.adsabs.harvard.edu/abs/2000ApJS..128..171S}{128, 171}

\bibitem[{{Snowden} {et~al.}(1991){Snowden}, {Mebold}, {Hirth}, {Herbstmeier}, \& {Schmitt}}]{Snowden1991Sci}
{Snowden}, S.~L., {Mebold}, U., {Hirth}, W., {Herbstmeier}, U., \& {Schmitt}, J.~H.~M. 1991, \href{http://dx.doi.org/10.1126/science.252.5012.1529}{\color{magenta}Science}, \href{https://ui.adsabs.harvard.edu/abs/1991Sci...252.1529S}{252, 1529}

\bibitem[{{Sofue} \& {Reich}(1979)}]{Sofue1979AA}
{Sofue}, Y. \& {Reich}, W. 1979, \aaps, \href{https://ui.adsabs.harvard.edu/abs/1979A&AS...38..251S}{38, 251}

\bibitem[{{Sunyaev} {et~al.}(2021){Sunyaev}, {Arefiev}, {Babyshkin}, {Bogomolov}, {Borisov}, {Buntov}, {Brunner}, {Burenin}, {Churazov}, {Coutinho}, {Eder}, {Eismont}, {Freyberg}, {Gilfanov}, {Gureyev}, {Hasinger}, {Khabibullin}, {Kolmykov}, {Komovkin}, {Krivonos}, {Lapshov}, {Levin}, {Lomakin}, {Lutovinov}, {Medvedev}, {Merloni}, {Mernik}, {Mikhailov}, {Molodtsov}, {Mzhelsky}, {M{\"u}ller}, {Nandra}, {Nazarov}, {Pavlinsky}, {Poghodin}, {Predehl}, {Robrade}, {Sazonov}, {Scheuerle}, {Shirshakov}, {Tkachenko}, \& {Voron}}]{Sunyaev2021AA}
{Sunyaev}, R., {Arefiev}, V., {Babyshkin}, V., {et~al.} 2021, \href{http://dx.doi.org/10.1051/0004-6361/202141179}{\color{magenta}\aap}, \href{https://ui.adsabs.harvard.edu/abs/2021A&A...656A.132S}{656, A132}

\bibitem[{{Truemper}(1982)}]{Truemper1982AdSpR}
{Truemper}, J. 1982, \href{http://dx.doi.org/10.1016/0273-1177(82)90070-9}{\color{magenta}Advances in Space Research}, \href{https://ui.adsabs.harvard.edu/abs/1982AdSpR...2d.241T}{2, 241}

\bibitem[{{Tumlinson} {et~al.}(2017){Tumlinson}, {Peeples}, \& {Werk}}]{Tumlinson2017ARAA}
{Tumlinson}, J., {Peeples}, M.~S., \& {Werk}, J.~K. 2017, \href{http://dx.doi.org/10.1146/annurev-astro-091916-055240}{\color{magenta}\araa}, \href{https://ui.adsabs.harvard.edu/abs/2017ARA&A..55..389T}{55, 389}

\bibitem[{{Wilms} {et~al.}(2000){Wilms}, {Allen}, \& {McCray}}]{Wilms2000ApJ}
{Wilms}, J., {Allen}, A., \& {McCray}, R. 2000, \href{http://dx.doi.org/10.1086/317016}{\color{magenta}\apj}, \href{https://ui.adsabs.harvard.edu/abs/2000ApJ...542..914W}{542, 914}

\bibitem[{{Yeung} {et~al.}(2023){Yeung}, {Freyberg}, {Ponti}, {Dennerl}, {Sasaki}, \& {Strong}}]{2023A&A...676A...3Y}
{Yeung}, M.~C.~H., {Freyberg}, M.~J., {Ponti}, G., {et~al.} 2023, \href{http://dx.doi.org/10.1051/0004-6361/202345867}{\color{magenta}\aap}, \href{https://ui.adsabs.harvard.edu/abs/2023A&A...676A...3Y}{676, A3}

\bibitem[{Zonca {et~al.}(2019)Zonca, Singer, Lenz, Reinecke, Rosset, Hivon, \& Gorski}]{HealpyZonca2019}
Zonca, A., Singer, L., Lenz, D., {et~al.} 2019, \href{http://dx.doi.org/10.21105/joss.01298}{\color{magenta}Journal of Open Source Software}, 4, 4

\end{thebibliography}

\begin{appendix}

\section{Correction tracking}

\begin{figure*}[b!]
    \centering
    \includegraphics[width=0.9\textwidth, trim=10 0 0 40,clip]{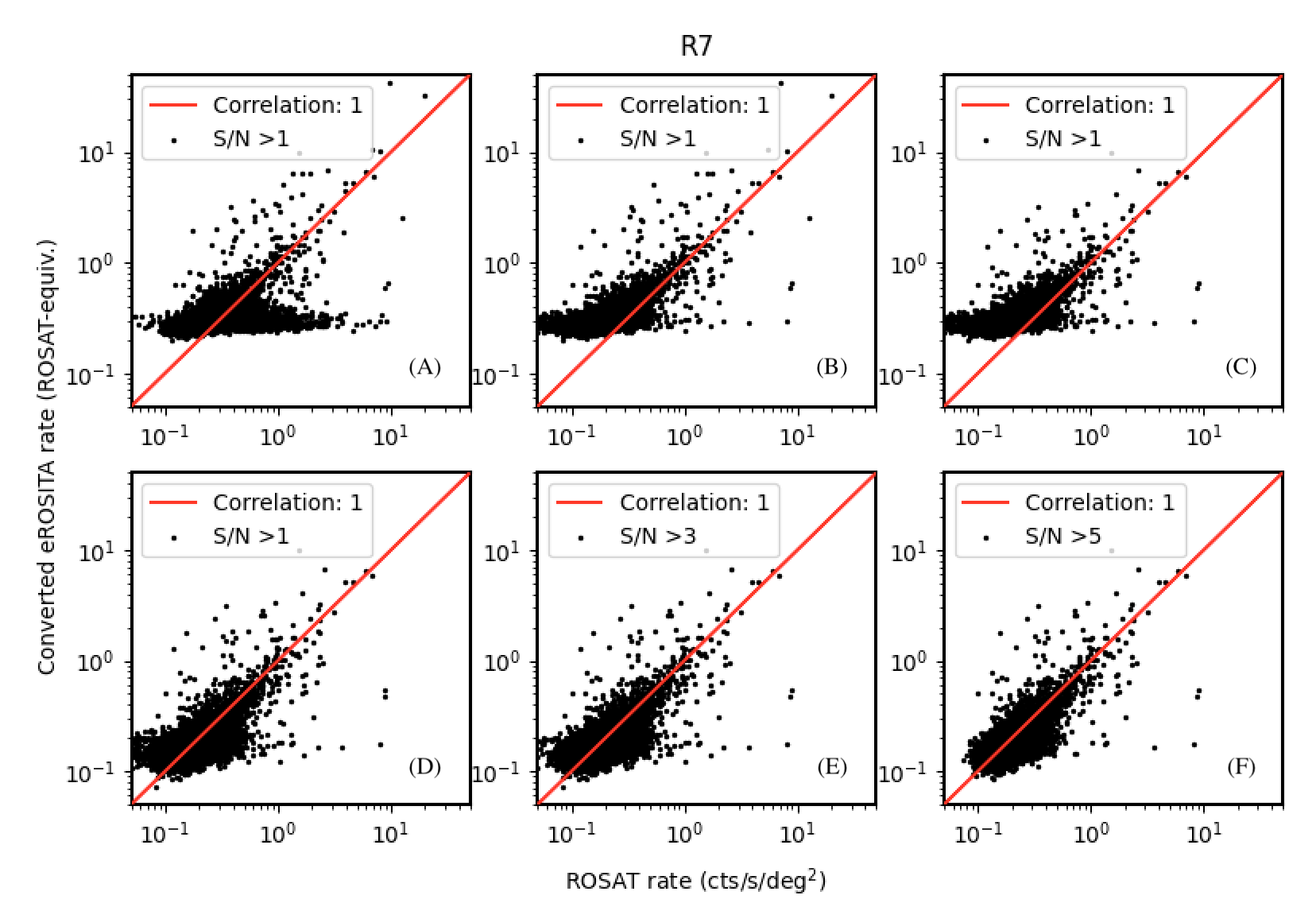}
    \caption{Detailed view of the \rosat{} and \erosita{} comparison in the R7 energy band. \small{(A)--(B): Horizontal branch deviation (shown in A) without the negative values. Instead, keeping and smoothing them results in the improved plot (B). This corresponds to the correction "ROSAT elliptic longitudinal stripes" in the text. 
    (B)--(C): We masked out the Sco X-1 in plot (C) because the single-reflection effect is stronger in \erosita{}. By this, we removed about 30 data points, which exceed than 1-1 relation as for the result. This corresponds to the correction in "Light reflection".
    (C)--(D): Particle background of \erosita{} was subtracted in this step to match the \rosat{} data, which are free of particle background.
    (D)--(E)--(F): Selection over \rosat{} S/N. Increasing the S/N threshold sharpens the 1-1 relation especially in fainter samples. This step corresponds to the "uncertainty selection."}} 
    \label{fig:correction1}
\end{figure*}

In Sec.\ref{sec:compare} we present different corrections that aimed to prepare the\erosita{} data for a quantitative comparison with the \rosat{} data, as presented in Fig.\ref{fig:4relations}. In Fig.\ref{fig:correction1} we provide versions of the data in with each correction is applied or is not applied.

\section{Projection distortion}

\begin{figure*}
\centering
\includegraphics[width=0.9\textwidth, trim=470 20 50 0,clip, angle=0]{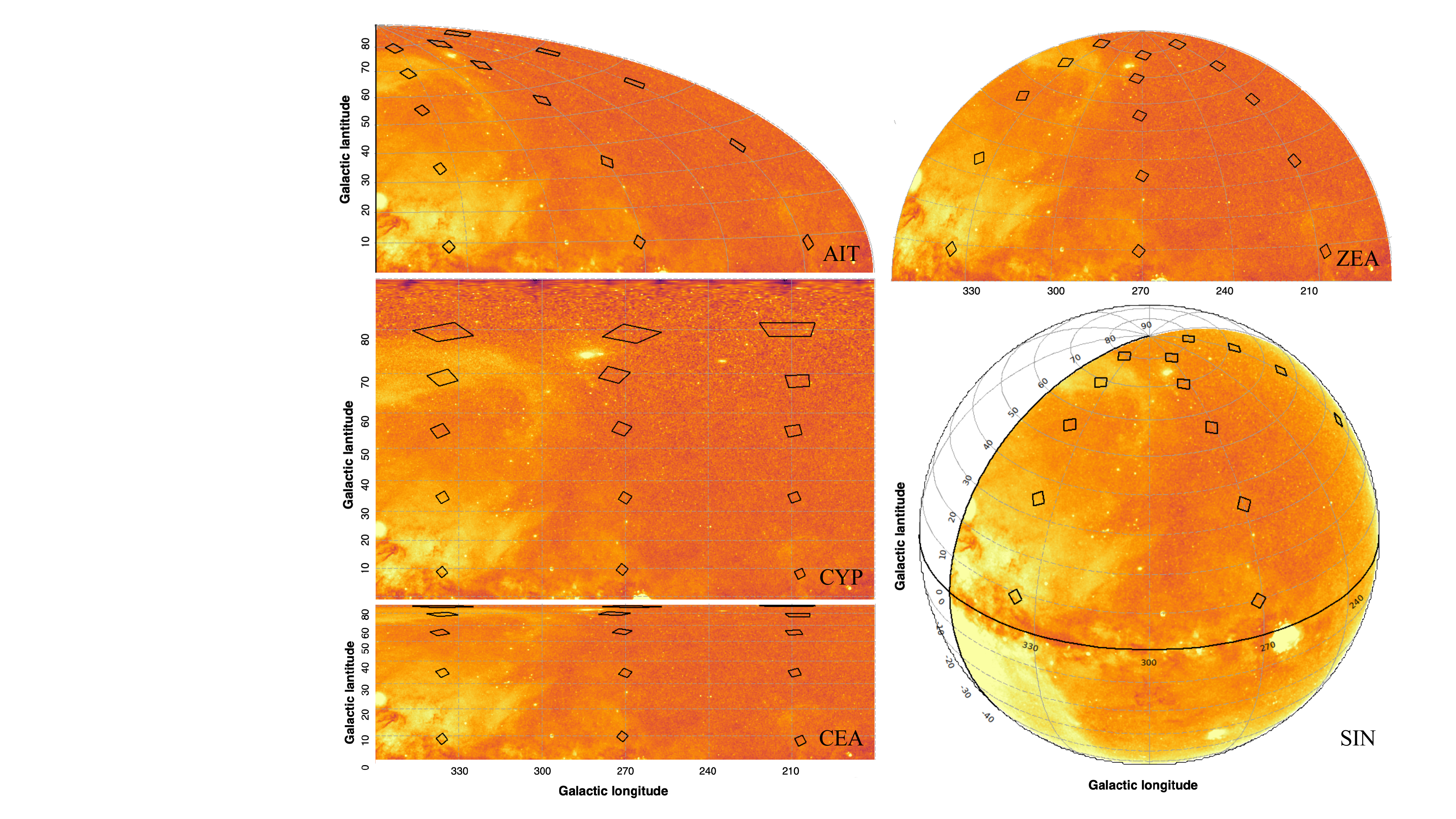}
\caption{\small{Distortions of the squared 3x3 $\rm degree^2$ sky tiles (in tangential projection around the center of the sky tile) when they are differently projected. The other projections represented here are CYP, CEA, AIT, ZEA, and orthographic (SIN) projections. Only the northwest Galactic hemisphere is shown.
In a nonequal area projection (e.g., CYP), the shape of a region is distorted and the projected area of each equal sky tile is not conserved.
The sky tiles at higher Galactic latitudes appear to be stretched. }} 
\label{fig:exp_distortion}
\end{figure*}

For different displaying purposes, we used different projections in this paper.
In Fig. \ref{fig:exp_distortion} we show the distortion effect of projections over eRASS1: 0.6--1.0 keV map. The cylindrical perspective (CYP) projection does not conserve the pixel area at different coordinates. The maps in CYP projection were corrected for the weight maps to recover the surface brightness.

\section{Determining the map resolution}

\begin{figure}[b!]
    \centering
    \includegraphics[width=0.48\textwidth, trim=0 0 0 30,clip]{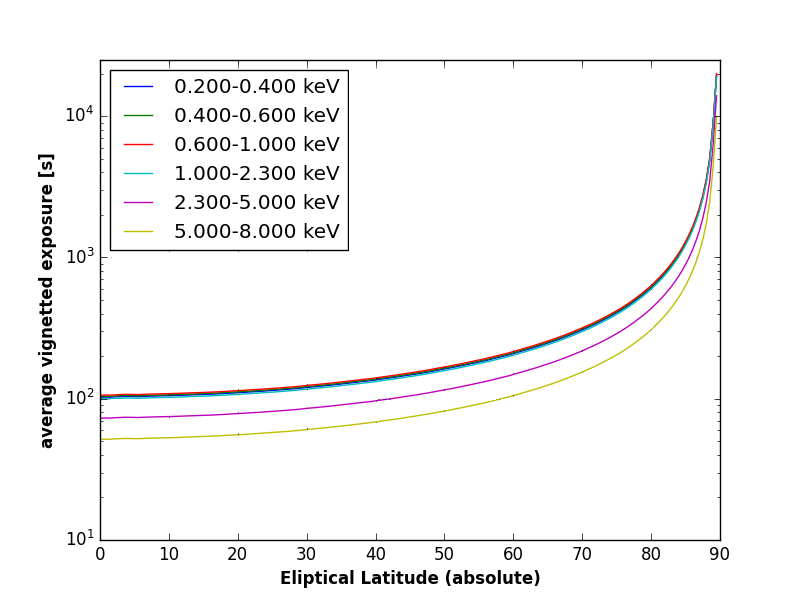}
    \caption{Average exposure (vignette corrected) per TM in different energy bands of eRASS1 as a function of ecliptic latitude.}
    \label{fig:exp_latitude} 
\end{figure}

\begin{figure}[t]
    \centering
    \includegraphics[width=0.425\textwidth, trim=0 0 0 10,clip]{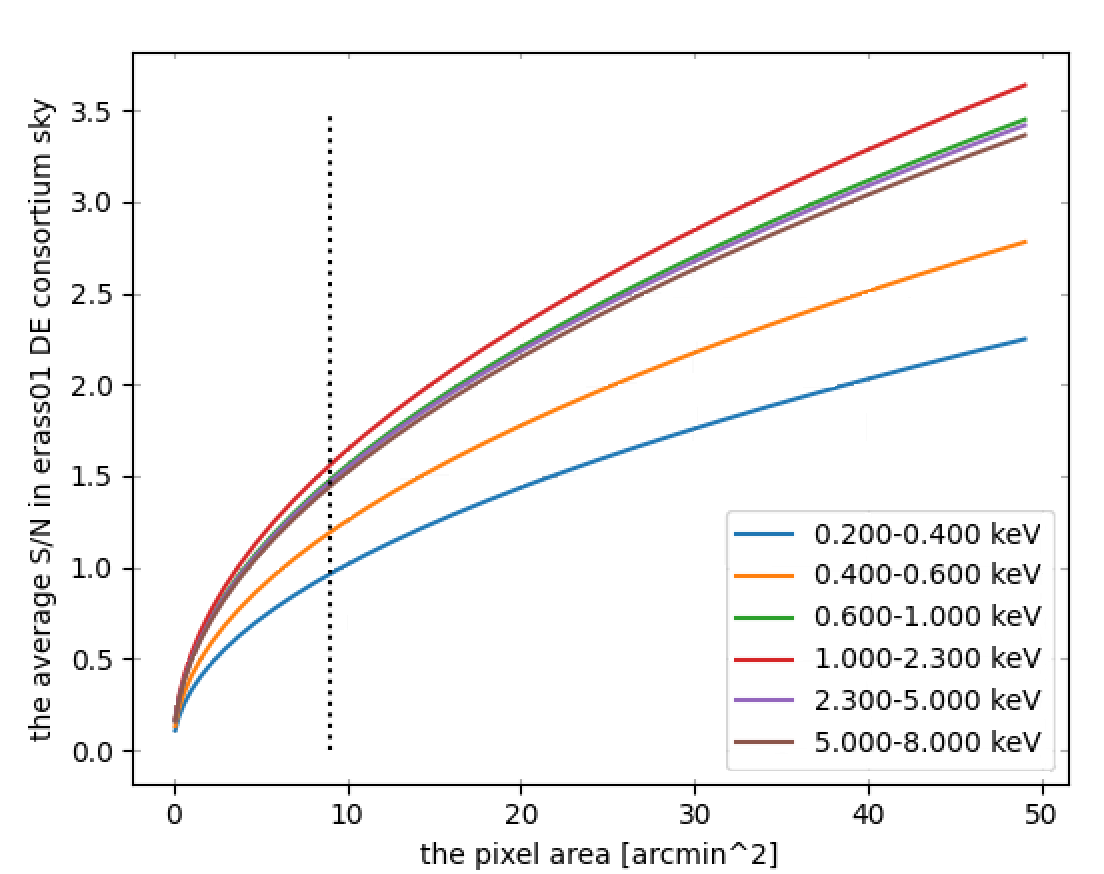}
    \caption{Average significance in different eRASS1 energy bands as a function of the pixel size. The vertical dashed line is set at 9 arcmin$^2$, used in this work.}
    \label{fig:significance} 
\end{figure}

The maps are presented in a pixel size of 3 arcmin.
This setup was chosen to balance the S/N in the bright and faint sky area.
Because of the scanning strategy, the exposure accumulates at the ecliptic poles (see the exposure, vignette corrected, at different energies in Fig. \ref{fig:exp_latitude}). According to the exposure, the significance of the detection is calculated in Fig. \ref{fig:significance}. The y-axis reports the average S/N over half the sky. The 9 arcmin$^2$ pixel satisfies the $S/N\geq 1$ in all the broadband maps. 

\section{Contribution of clusters in eRASS1}

\begin{figure*}[h]
    \centering
    \includegraphics[width=0.90\textwidth, trim=40 0 30 0,clip]{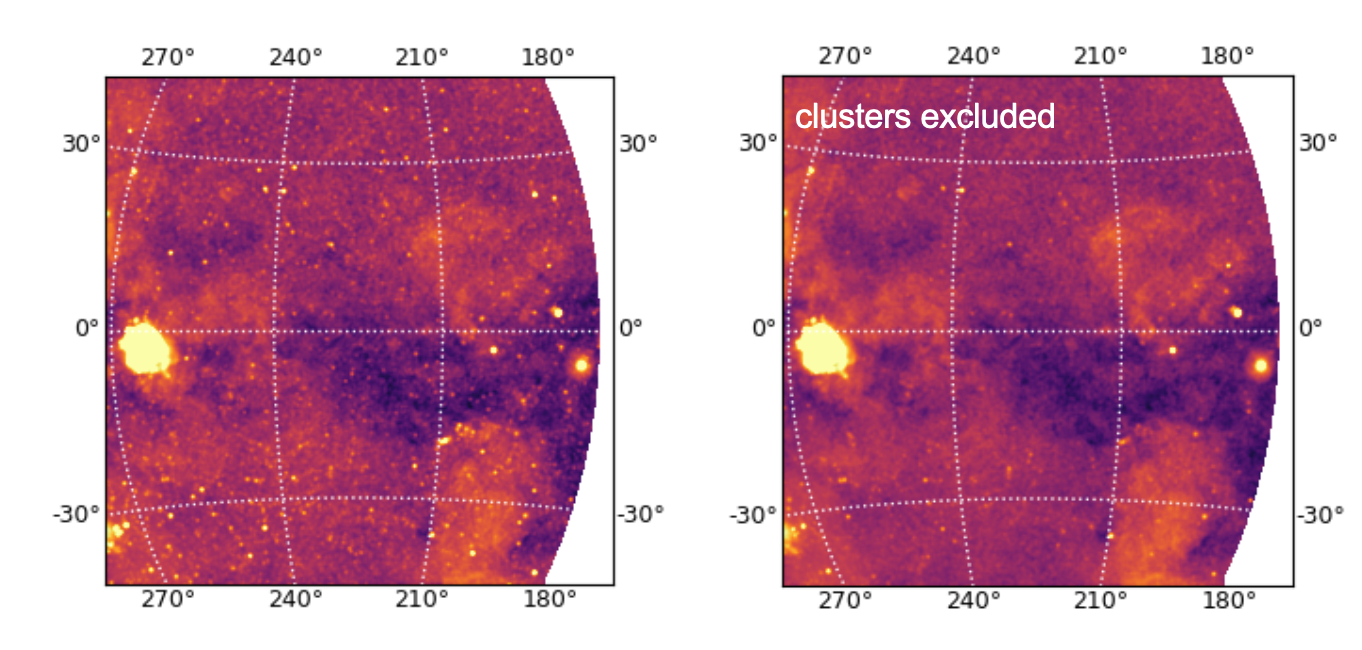}
    \caption{\small{eRASS1 0.44-1.01 keV band with and without galaxy cluster candidates (left and right). Point sources were subtracted from both maps as described in Sec.\ref{sec:subtract}. The cluster candidates are extracted from the extended sources presented in the eRASS1 catalog \citep{Merloni23}}}
    \label{fig:cluster}
\end{figure*}

The sensitivity of \erosita{} enables us to probe the sky to a larger depth. Hence, faint extended sources such as galaxy clusters can be distinguished. As the main scientific goal of \erosita{}, the detection of galaxy clusters candidates is expected to extend their census by more than one order of magnitude. Fig.\ref{fig:cluster} (point sources subtracted) shows a still considerable number of radiating objects throughout the sky, even in the Galactic plane. Most of them are galaxy clusters that are classified in an upcoming eRASS1 cluster catalog. In contrast to \erosita{}, \rosat{} does not observe the faintest end of the cluster luminosity function (i.e., the largest amount), due to small count statistics. In the field we show here, $\sim$ 300 clusters candidates are present. After further subtraction of these extended sources, a smoother large-scale diffuse emission can be obtained.

\section{High-resolution \erosita{} eRASS1 broadband maps}

Here we present the the gallery of high-resolution (pixel size of 3 arcmin) eRASS1 broadband maps of bands 0.2--0.4 keV, 0.4--0.6 keV, 0.6--1.0 keV, 1.0--2.3 keV, and 2.3--5.0 keV. The final maps are available in fits and HealPix format. More details of the map processing can be found in Sec.\ref{sec:process}.

\begin{figure*}[htbp!]
\centering
\includegraphics[width=\textwidth, trim=0 45 10 45, clip]{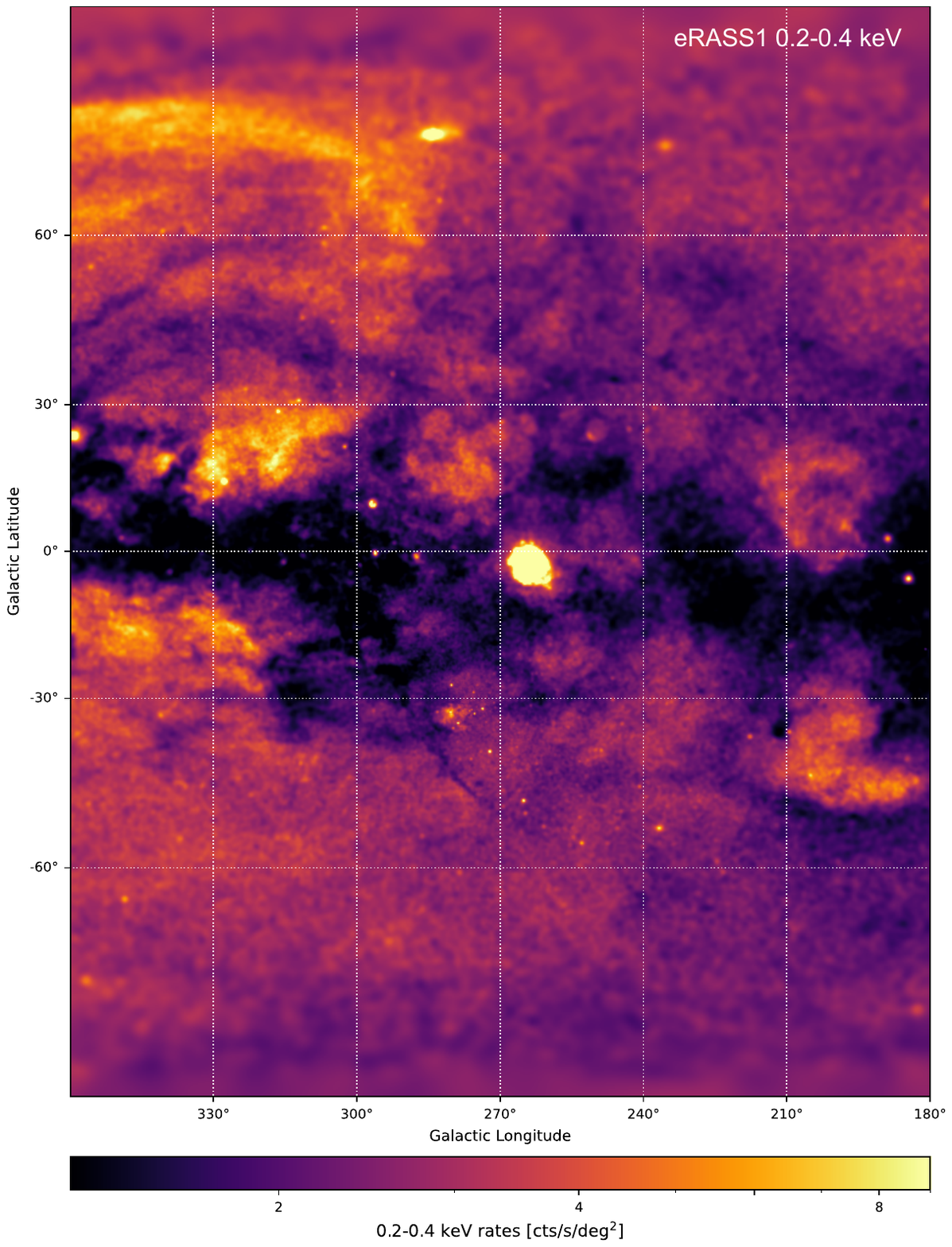}
\caption{Broadband 0.2--0.4~keV eRASS1 map in CYP. The color bar shows the range $\rm 1.0-9.0\, cts\, s^{-1}\, deg^{-2}$ in log scale. An adaptive smoothing with S/N $\ge$ 20 is used. The minimum threshold is set at the instrumental background of this energy band.}
\label{fig:0.2_0.4}
\end{figure*}

\begin{figure*}[htbp!]
\centering
\includegraphics[width=\textwidth,trim=0 45 10 45, clip]{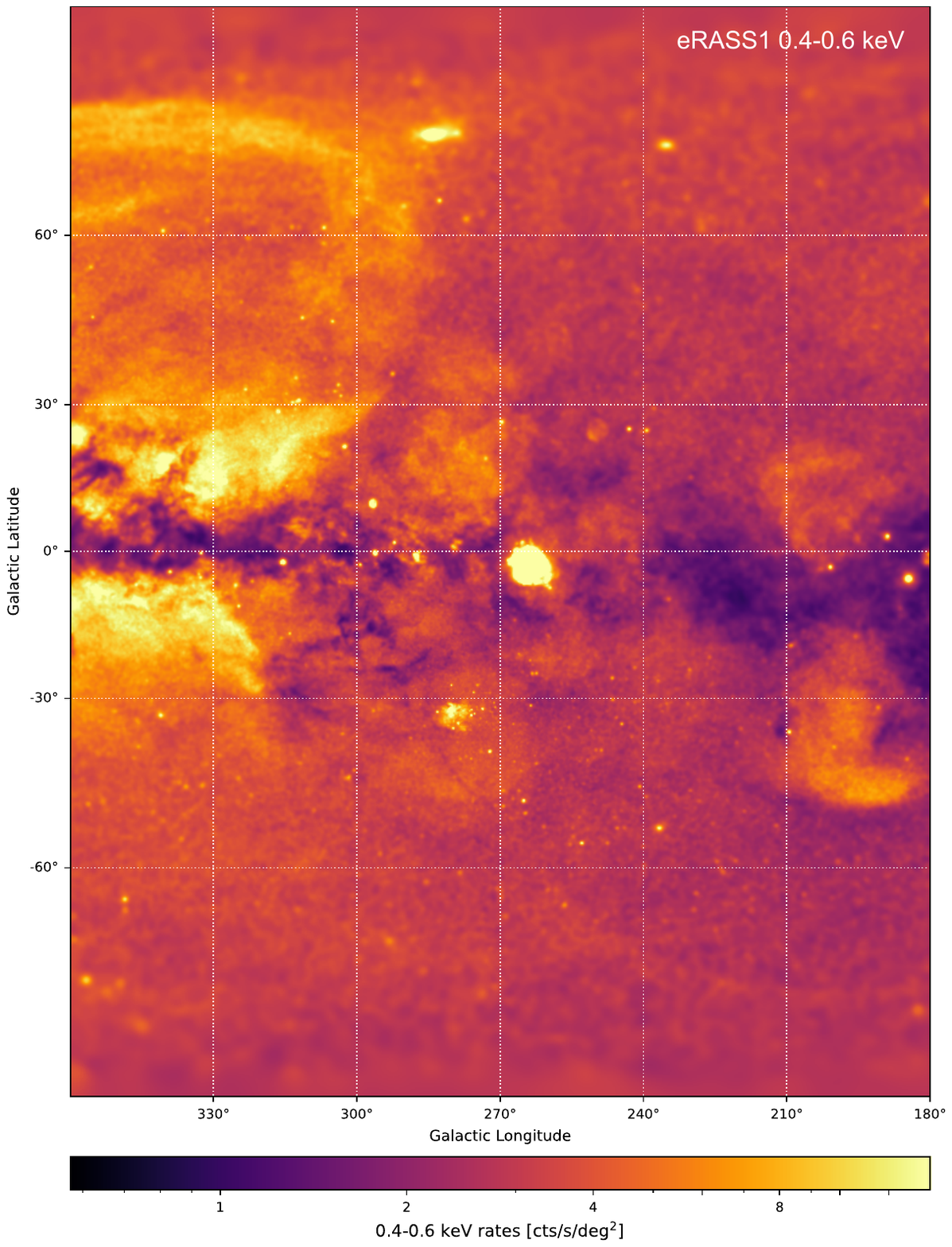}
\caption{Broadband 0.4--0.6~keV eRASS1 map in CYP. The color bar shows the range $\rm 1.0-14.0\, cts\, s^{-1}\, deg^{-2}$ in log scale. An adaptive smoothing with S/N $\ge$ 20 is used. The minimum threshold is set at the instrumental background of this energy band.}
\label{fig:0.4_0.6}
\end{figure*}

\begin{figure*}[htbp!]
\centering
\includegraphics[width=\textwidth,trim=0 45 10 45, clip]{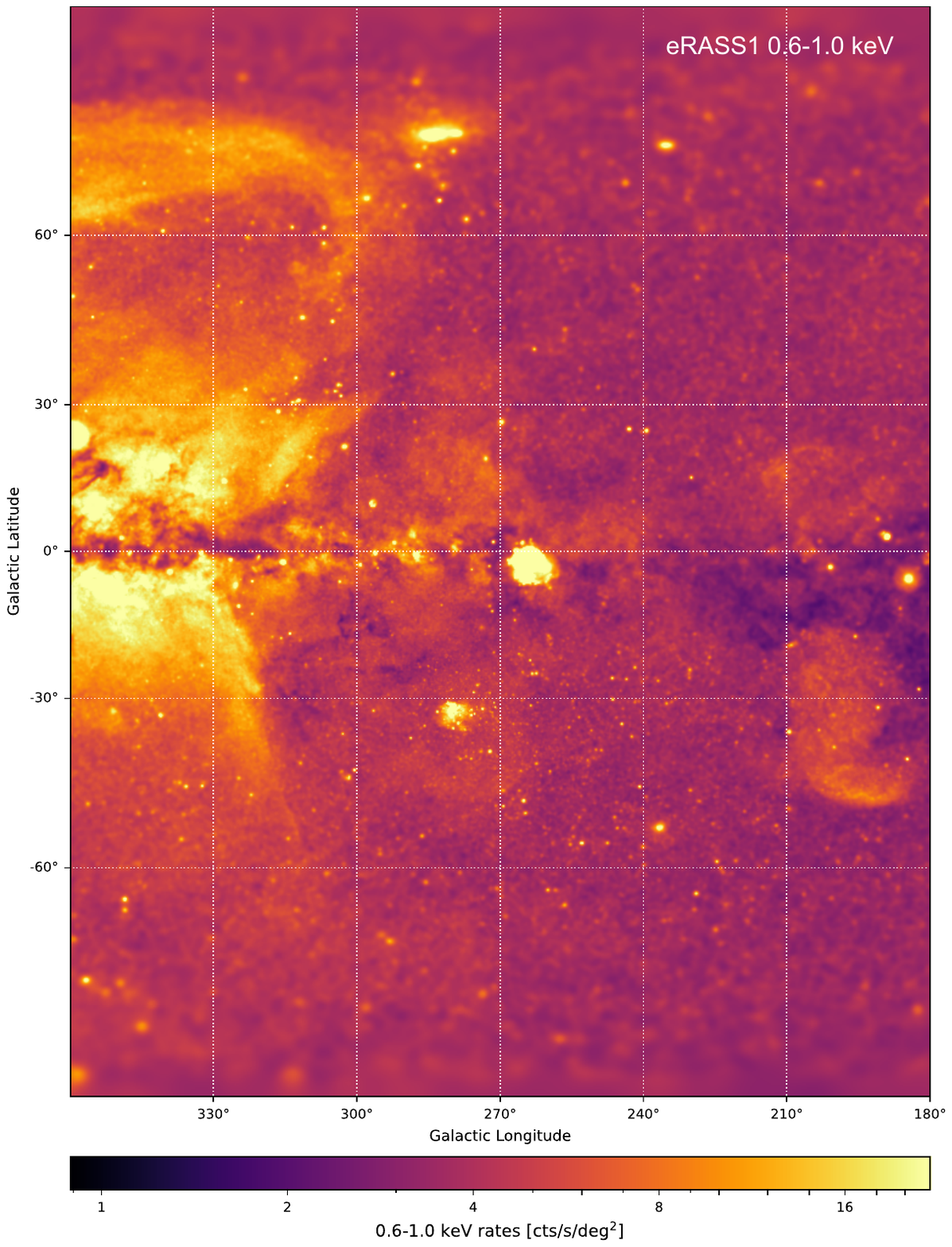}
\caption{Broadband 0.6--1.0~keV eRASS1 map in CYP. The color bar shows the range $\rm 1.0-22.0\, cts\, s^{-1}\, deg^{-2}$ in log scale. An adaptive smoothing with S/N $\ge$ 20 is used. The minimum threshold is set at the instrumental background of this energy band.}
\label{fig:0.6_1.0}
\end{figure*}

\begin{figure*}[htbp!]
\centering
\includegraphics[width=\textwidth,trim=0 45 10 45, clip]{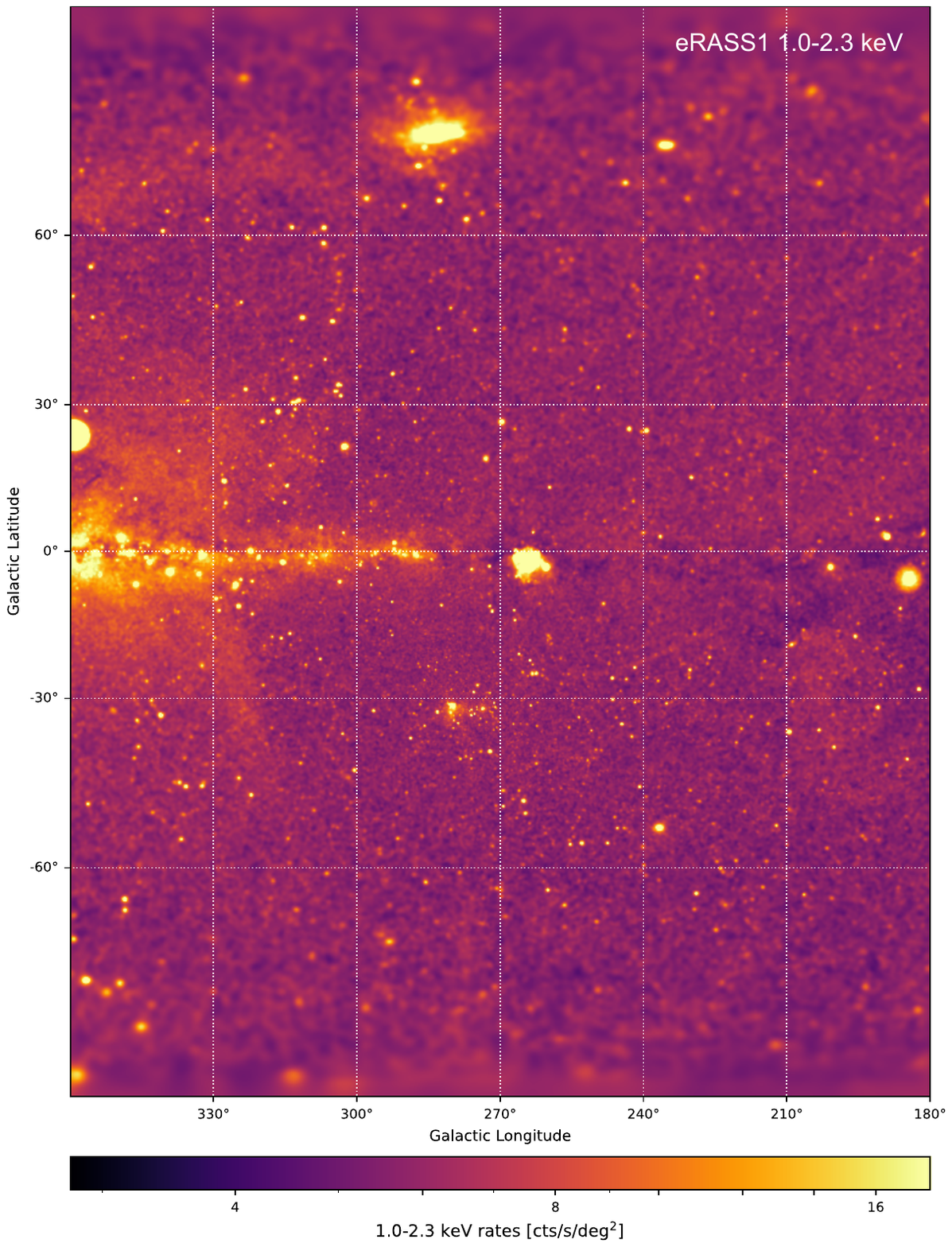}
\caption{Broadband 1.0--2.3~keV eRASS1 map in CYP. The color bar shows the range $\rm 2.5-18.0\, cts\, s^{-1}\, deg^{-2}$ in log scale. An adaptive smoothing with S/N $\ge$ 20 is used. The minimum threshold is set at the instrumental background of this energy band.}
\label{fig:1.0_2.3}
\end{figure*}

\begin{figure*}[htbp!]
\centering
\includegraphics[width=\textwidth,trim=0 45 10 45, clip]{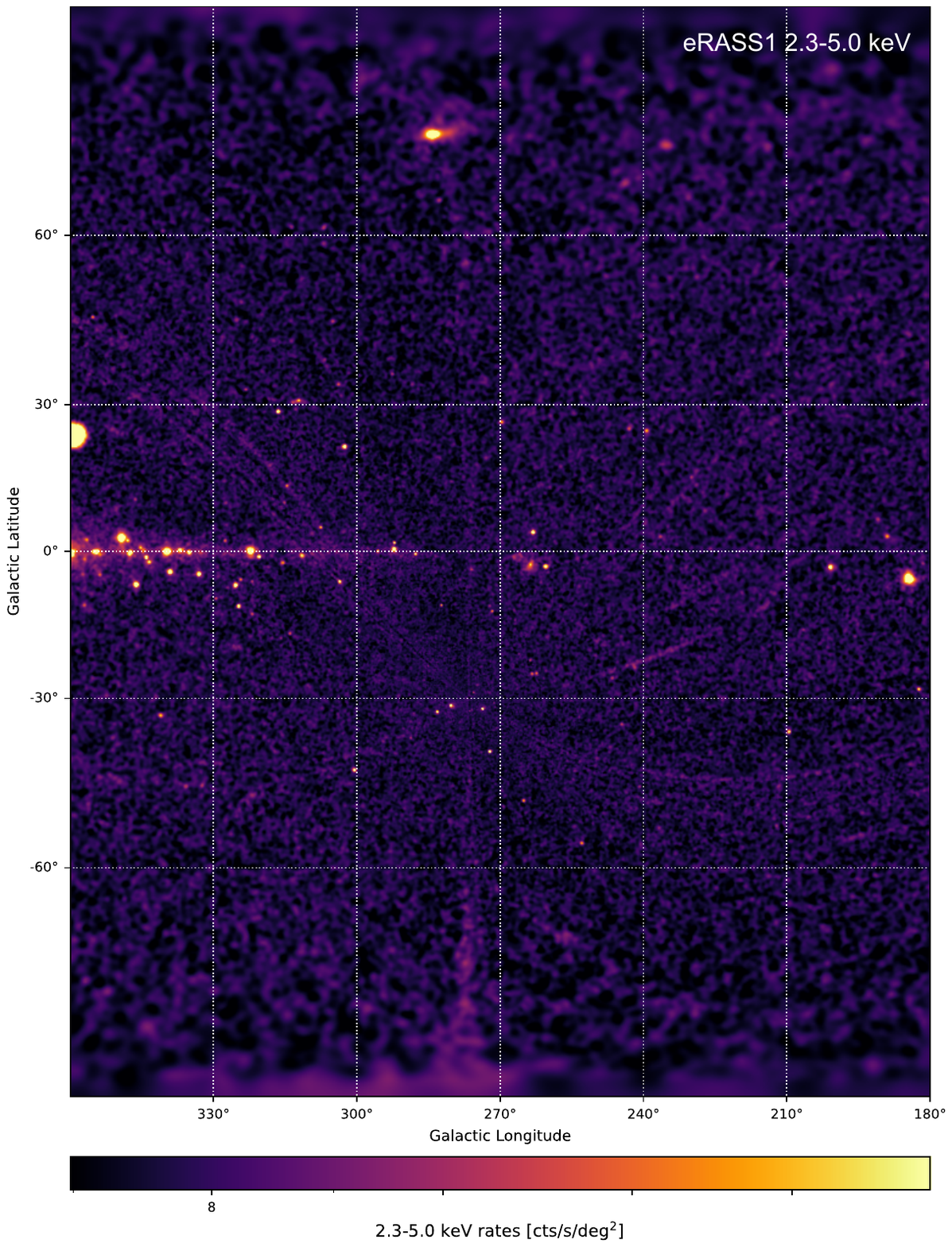}
\caption{Broadband 2.3--5.0~keV eRASS1 map in CYP. The color bar shows the range $\rm 5.5-13.0\, cts\, s^{-1}\, deg^{-2}$ in log scale. An adaptive smoothing with S/N $\ge$ 20 is used. The minimum threshold is set at the instrumental background of this energy band.}
\label{fig:2.3_5.0}
\end{figure*}

\section{Additional maps of \erosita{} and \rosat{}}

Additional plots are shown here for a detailed comparison of the \erosita{} and \rosat{} count rates. By placing the sextant section of the eRASS1 and RASS maps in parallel, the comparison can be verified by direct visual inspection.

\begin{figure*}
\centering
\includegraphics[width=\textwidth,trim=10 53 10 33, clip]{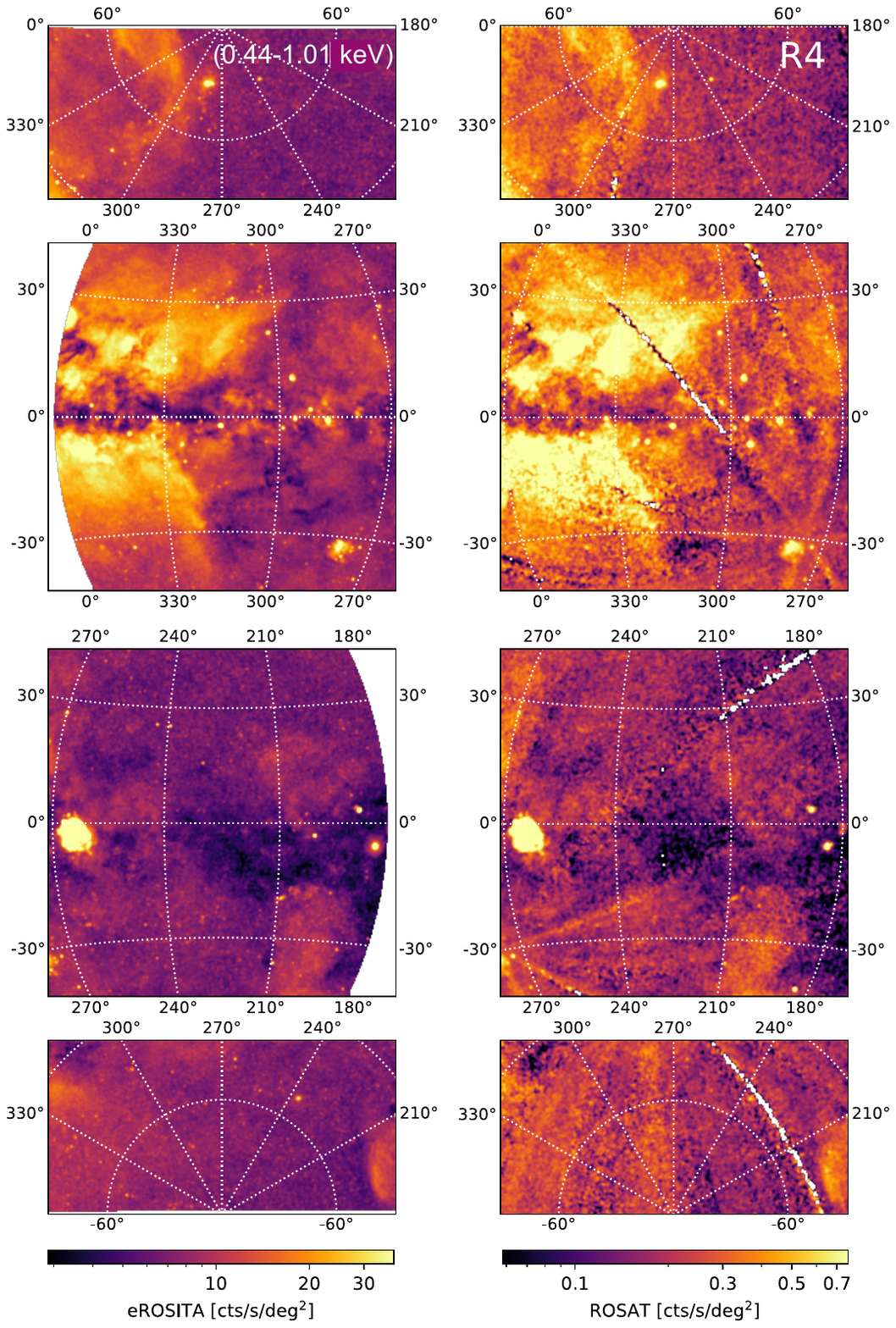}
\caption{Count-rate maps of \erosita{} (left column) and \rosat{} (right column) in the R4 band (0.44--1.01 keV). A ZEA projection is used. North is up. The log-scaled color bar covers 90\% of the dynamic pixel range.}
\label{fig:ero_r4}
\end{figure*}

\begin{figure*}
\centering
\includegraphics[width=\textwidth,trim=10 53 10 33, clip]{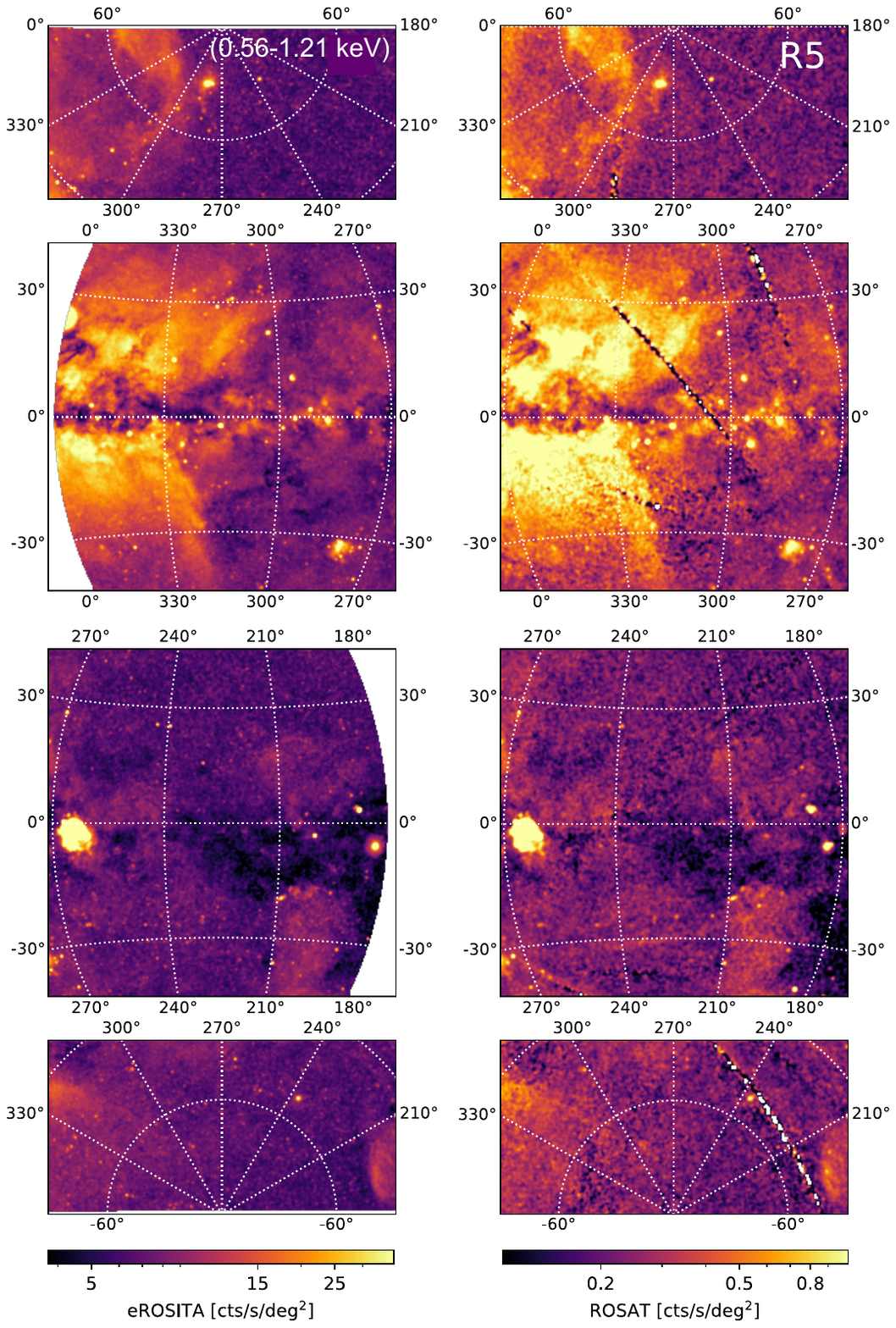}
\caption{Count-rate maps of \erosita{} (left column) and \rosat{} (right column) in the R5 band (0.56--1.21 keV). A ZEA projection is used. North is up. The log-scaled color bar covers 90\% of the dynamic pixel range.}
\label{fig:ero_r5}
\end{figure*}

\begin{figure*}
\centering
\includegraphics[width=\textwidth,trim=10 53 10 33, clip]{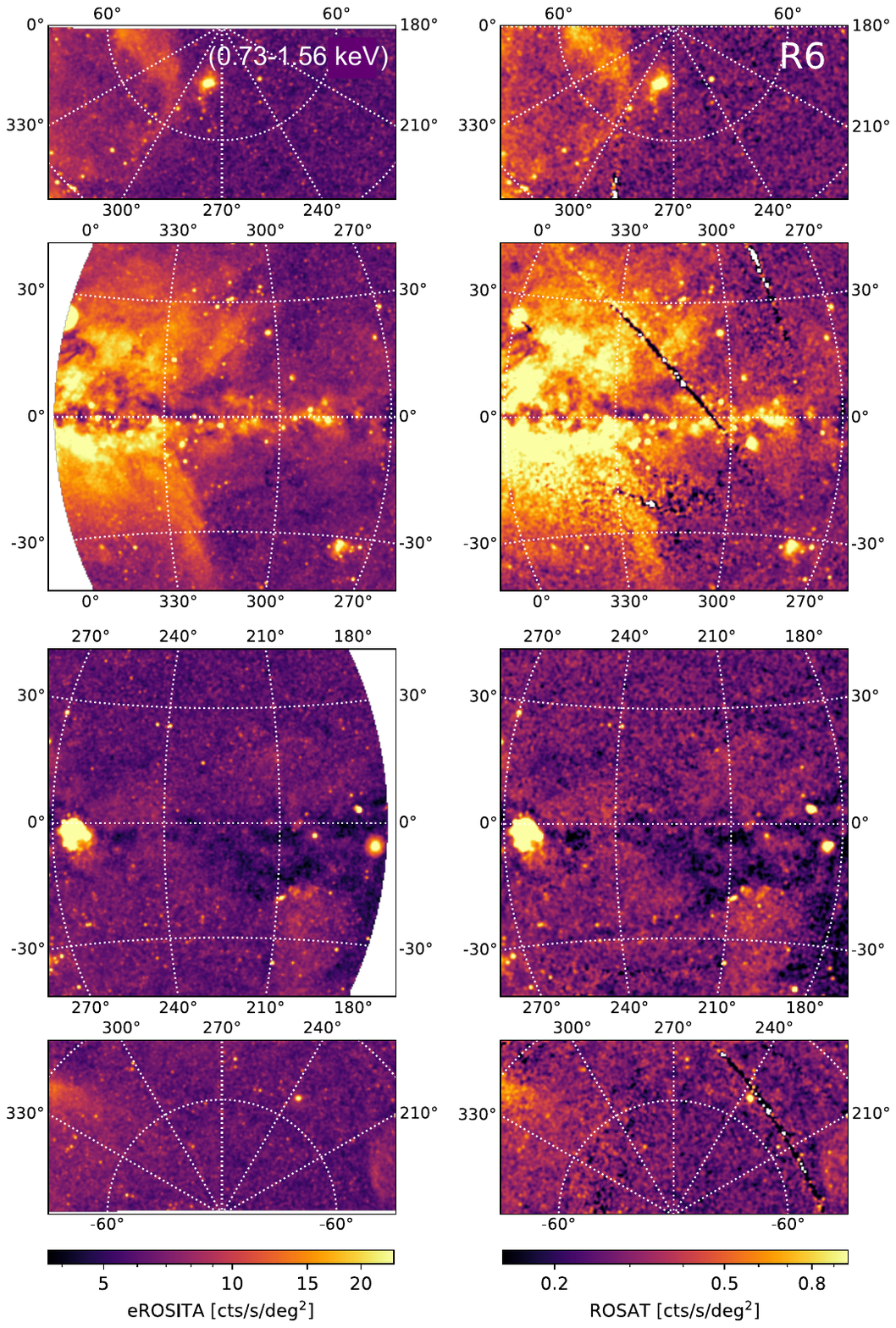}
\caption{Count-rate maps of \erosita{} (left column) and \rosat{} (right column) in the R6 band (0.73--1.56 keV). A ZEA projection is used. North is up. The log-scaled color bar covers 90\% of the dynamic pixel range.}
\label{fig:ero_r6}
\vspace{-10pt}
\end{figure*}

\begin{figure*}
\centering
\includegraphics[width=\textwidth,trim=10 53 10 33, clip]{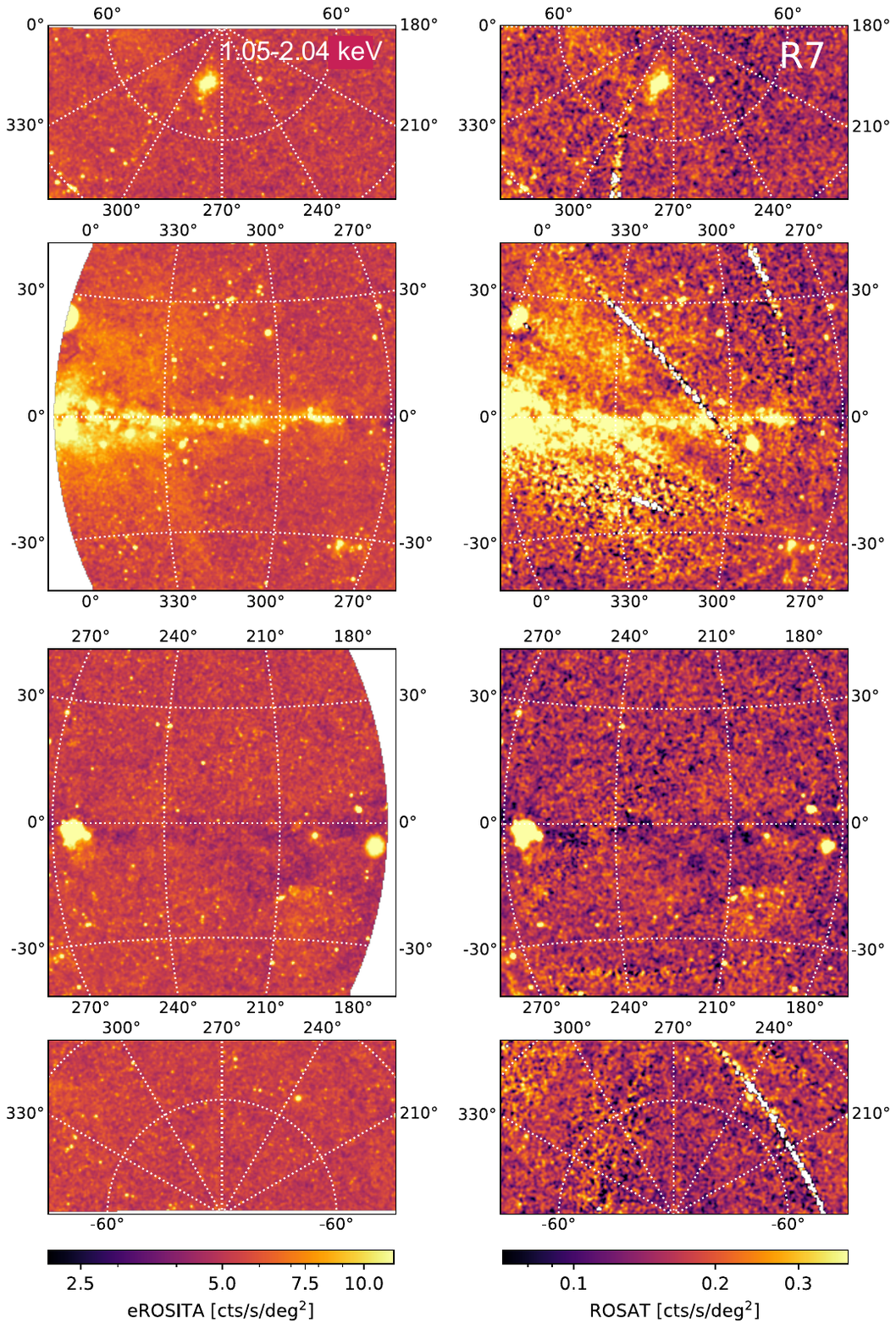}
\caption{Count-rate maps of \erosita{} (left column) and \rosat{} (right column) in the R7 band (1.05--2.04 keV). A ZEA projection is used. North is up. The log-scaled color bar covers 90\% of the dynamic pixel range.}
\label{fig:ero_r7}
\end{figure*}

\end{appendix}
\end{document}